\documentclass[twocolumn,superscriptaddress,prb,floatfix]{revtex4}
\usepackage{graphicx,amsfonts,amsmath,latexsym,amssymb}

\begin{document}

\title{
Magnetic analogue of the LOFF phase in Sr$_3$Ru$_2$O$_7$}

\author{A. M. Berridge}
\affiliation{School of Physics and Astronomy, University of Birmingham,
Edgbaston, Birmingham B15 2TT, UK}

\affiliation{School of Physics and Astronomy, University of St Andrews,
North Haugh, St Andrews KY16\ 9SS, UK}

\author{S. A. Grigera}
\affiliation{School of Physics and Astronomy, University of St Andrews,
North Haugh, St Andrews KY16\ 9SS, UK}

\affiliation{Instituto de F\'{\i}sica de L\'{\i}quidos y Sistemas
  Biol\'ogicos, UNLP, La Plata 1900, Argentina} 

\author{B. D. Simons}
\affiliation{Cavendish Laboratory, University of Cambridge, Madingley Road,
Cambridge, CB3\ 0HE, UK}

\author{A. G. Green}
\affiliation{School of Physics and Astronomy, University of St Andrews,
North Haugh, St Andrews KY16\ 9SS, UK}

\date{\today}

\begin{abstract}
The phase diagram of Sr$_3$Ru$_2$O$_7$ contains a metamagnetic transition that bifurcates to enclose an anomalous phase with intriguing properties - a large resistivity with anisotropy that breaks the crystal-lattice symmetry. We propose that this is a magnetic analogue of the spatially inhomogeneous superconducting Fulde-Ferrell-Larkin-Ovchinnikov state.  Based on a microscopic theory of Stoner magnetism we derive a Ginzburg-Landau expansion where the magnetisation transverse to the applied field can become spatially inhomogeneous.  We show that this reproduces the observed phase diagram of Sr$_3$Ru$_2$O$_7$.
\end{abstract}

\maketitle
\section{Introduction}

Phase transitions may occur through the sudden or continuous appearence of order.  A third possibility, that the phase is established via an intermediate state with spatially modulated order was postulated by Fulde and Ferrell~\cite{Fulde}, and Larkin and Ovchinnikov~\cite{Larkin}, in the case of superconductivity.  They showed that a spatially modulated superconducting phase is favourable in a narrow range of magnetic fields between the normal and superconducting states.  Despite stimulating great interest since its proposal the Fulde-Ferrell-Larkin-Ovchinnikov (LOFF) phase has yet to be conclusively observed.  It is however the source of much interest in cold atoms and heavy-fermion and color superconductivity~\cite{Loff_review}.  These ideas also have links to the search for an intermediate phase between the isotropic Fermi liquid and the Wigner crystal where the electronic liquid breaks some but not all of the symmetries of its environment~\cite{Fradkin07}.  There is a magnetic analogue to these phenomena.  The formation of a spatially modulated magnetization may pre-empt the transition between uniform paramagnet and ferromagnet, or a metamagnetic transition.

Such a mechanism may explain the puzzling properties of Sr$_3$Ru$_2$O$_7$~\cite{Grigera01,Grigera03a,Grigera04,Borzi07,Mercure,Rost}.  This material shows a complex phase diagram where a metamagnetic transition may be tuned by varying the angle of the applied magnetic field with respect to the crystal axes.  As the critical endpoint of this first order phase transition approaches zero temperature the transition bifurcates to enclose a region with striking transport properties.  The resistivity in this region is anomalously high and shows anisotropy.  When current is passed in the direction most perpendicular to the applied field then the resistivity drops away rapidly if the field is applied at an angle from the c-axis.  With the current in the direction most parallel to the field then the high resistivity persists for a greater range of angle.  We propose that this anomalous region is a phase of spatially modulated magnetisation appearing between the low- and high-field sides of the metamagnetic transition, in analogy with the superconducting LOFF state.

This article is an extension of the ideas we presented in Ref. \onlinecite{Berridge}.  We begin by describing heuristically why modulated states may become favourable before turning explicitly to the metamagnetic system.  In a manner similar to Ref. \onlinecite{Binz04}, we show how a Stoner model with a peak in the electronic density of states can reproduce the metamagnetic transition.  We then consider how this can be extended to include modulated states.  We will show how in an expansion about the line of metamagnetic critical endpoints the transverse spin stiffness vanishes.  Based on this we will study a phenomenology for the modulated metamagnetic system and calculate how the metamagnetic transition reconstructs to accomodate the inhomogeneous phase.  We compare this model with the data on Sr$_3$Ru$_2$O$_7$.

\section{Heuristic Picture}

We begin with a heuristic discussion of inhomogeneous phase formation.  We concentrate on superconductivity as it was here that the intermediate modulated phase was first proposed.  We describe how analogous principles can apply to the formation of spatially modulated magnetic states.

A BCS superconductor is formed when electrons with spin-up and momentum $+{\bf k}$ pair with electrons with spin-down and momentum $-{\bf k}$ to form a Cooper pair with zero total momentum.  When a magnetic field is applied the Zeeman energy favours splitting the spin-up and -down Fermi surfaces, therefore breaking the pairing.  The Fulde-Ferrell-Larkin-Ovchinnikov~\cite{Fulde,Larkin} state occurs in a narrow range of fields where the system can take advantage of both Zeeman and pairing energies by pairing electrons with $+{\bf k}+\frac{{\bf q}}{2}$ and $-{\bf k}+\frac{{\bf q}}{2}$.  This produces Cooper pairs with non-zero total momentum, corresponding to a spatial modulation of the superconducting order with wavevector ${\bf q}$.  In this way the transition from superconducting to normal phase may occur via an intermediate modulated phase.  A range of possible superconducting textures are possible, made by superposing several modulation wavevectors~\cite{Loff_review}.  Which texture is most favourable depends on microscopic details.

Similar ideas find application in magnetic systems.  It may be favourable for modulated magnetic states to form as intermediate phases in metamagnetic transitions.  For magnetism to be favourable there must be an energetic advantage to creating an imbalance in the number of up- and down-spin electrons.  In a Stoner model this is provided by Coulomb interaction energy, which is balanced against a gain in single-particle kinetic energy due to the change in electron momentum necessary to create the imbalance.  When the density of states (DoS) at the Fermi surface is high enough the interaction energy wins and the system becomes ferromagnetic.  If there is a peak in the DoS then one of the spin species' Fermi surface can be tuned into the region of high DoS by the application of a magnetic field.  When this happens it will become favourable for the system to magnetize and a metamagnetic transition occurs~\cite{Wohlfarth,Shimizu,Binz04}.  This peak may be caused by, for example, a van Hove singularity in the electronic band dispersion.

\begin{figure}
\centering
\includegraphics[width=3in]{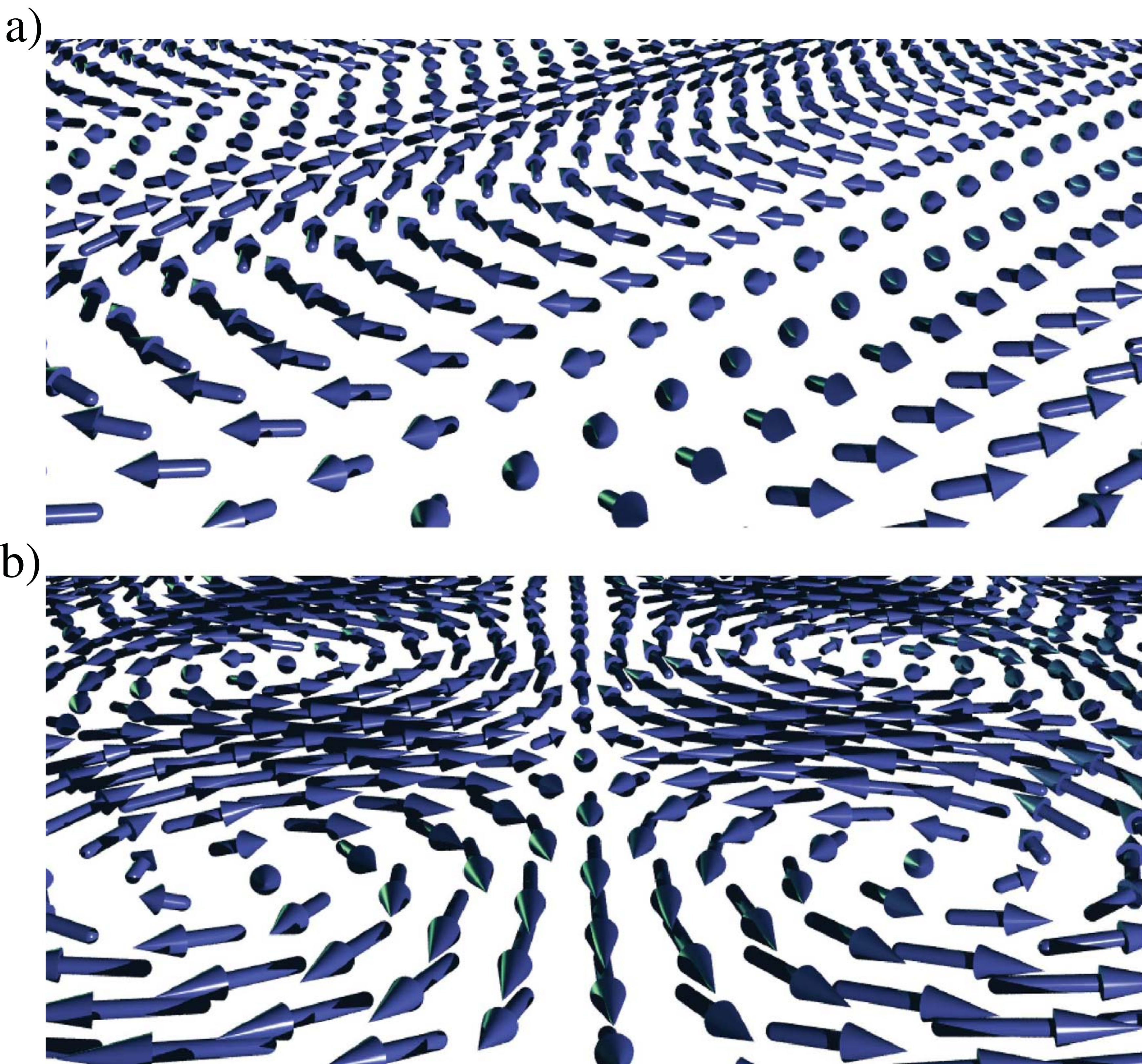}
\caption{\label{fig:texture} (Color online)  Possible magnetic textures.  a) A single spin-spiral.  b) A superposition of four spirals.  The longitudinal magnetic component has been suppressed for clarity.}
\end{figure}

Spatial modulation may be stabilized by similar considerations~\cite{Rice,Monthoux}.  The formation of modulation reconstructs the electronic dispersion, producing additional peaks in the DoS due to anticrossing of the electronic bands.  Advantage can be taken of these peaks in the same way as in the homogeneous case.  The single-particle energy cost to magnetising is reduced by occupying states under the new peak, leading to the favourability of modulated magnetisation.  As with the LOFF case a variety of spin-textures can be formed by superposing several wavevectors of modulation.  The possible textures will be determined by the symmetry of the band structure and may be sensitively dependent on microscopic details.  In this paper we will consider a single-${\bf q}$ state, the magnetic analogue of the Fulde-Ferrell ansatz.  A visulisation of a single-{\bf q} spiral and a spin-texture made by superimposing four such spirals, as may occur in a fourfold symmetric band structure, is presented in Fig.\ref{fig:texture}.

The analogy between superconductivity and magnetism is general~\cite{Schofield1,Schofield2}.  Magnetism can be viewed as pairing in the particle-hole, rather than particle-particle channel as in the superconducting case.  The mapping between the phase of the order parameter in the magnetic spin-spiral and the superconducting LOFF cases has also been made in the mapping of the XY model to superconductivity~\cite{XY}.

\section{Microscopic Model}

\subsection{The Stoner model}

Having discussed inhomogeneous phase formation generally, we now turn to the microscopic model which we will study.  We introduce the energetic reasons for the metamagnetic transition occuring and the methods by which the metamagnetic phase diagram may be calculated.  We discuss how this relates to Sr$_3$Ru$_2$O$_7$.

The dominant feature in the magnetic phase diagram of Sr$_3$Ru$_2$O$_7$ is the metamagnetic transition.  We begin from the idea that the full phase diagram, including the anomalous phase, is found by restructuring this metamagnetic transition.  Based on this idea we use the simplest model which reproduces the metamagnetic transition; the Stoner model with a peak in the electronic DoS.  This gives the correct topology for the metamagnetic transition with a generic 2D band dispersion~\cite{Binz04}.

We begin from the Hamiltonian:
\begin{eqnarray}
\hat{H}
-
\mu \hat{N}
&=&
\sum_{\bf k} \psi^\dagger_{\bf k}\left(\epsilon_{\bf k}-\mu\right)\psi_{\bf k}
-
\frac{g}{4} \int d{\bf x}\left(\psi^\dagger_{\bf x} {\boldsymbol \sigma} \psi_{\bf x}\right)^2
\nonumber\\
&&
-h \int d{\bf x}~ \psi^\dagger_{\bf x} \sigma_z \psi_{\bf x},
\label{Ham}
\end{eqnarray}
where $\psi^\dagger_{{\bf k}/{\bf x}}=\left(c^\dagger_{{\bf k}/{\bf x}, \uparrow}, c^\dagger_{{\bf k}/{\bf x}, \downarrow}\right)$ represents the electron creation operators in the momentum and position representations respectively, $\epsilon_{\bf k}$ is the electonic band dispersion, ${\boldsymbol \sigma}$ is the vector of Pauli matrices, $g$ is a contact interaction representing the screened Coulomb interaction and where we have singled out $z$ as the direction of the applied magnetic field $h$.  Here, the first term is the single-particle energy, the second term is the interaction energy and the final term is the Zeeman energy.  The balance between these terms will determine the magnetic properties of the system.  

We will consider the simplest dispersion which will produce a peak in the DoS, the 2D next-nearest-neighbour tight-binding dispersion $\epsilon_{\bf k}=-\left(\cos{k_x}+\cos{k_y}\right)+t \cos{k_x}\cos{k_y}$.  Here $t$ parameterises the amount of next-nearest neighbour hopping and prevents the pathological perfect nesting of the Fermi surface which causes the nearest-neighbour dispersion to become an antiferromagnetic insulator at van Hove filling.  This dispersion has saddle points at ${\bf k}=\left(0, \pm\pi\right)$ and ${\bf k}=\left(\pm\pi, 0\right)$.  These saddle points produce logarithmically divergent peaks in the DoS which will drive the metamagnetism~\cite{vanHove}.  The phenomenology which we develop is generic and not tied to this particular dispersion, relying only on a peak in the DoS near to the Fermi energy.  Details such as the wavevector of the inhomogeneity will depend on the exact dispersion used.

The Hamiltonian (\ref{Ham}) gives the well-known Stoner criterion for the formation of ferromagnetism, $g\rho_{\rm F}=1$ where $\rho_{\rm F}$ is the DoS at the Fermi surface.  This condition determines when a system will undergo a continuous transition between paramagnetic and ferromagnetic states.  When the DoS at the Fermi surface becomes large enough the system magnetizes.  By tuning the Fermi-surface through a peak in the DoS the system can be tuned through the ferromagnetic transition.

\subsubsection{Cartoon of the metamagnetic transition}

\begin{figure}
\centering
\includegraphics[width=3in]{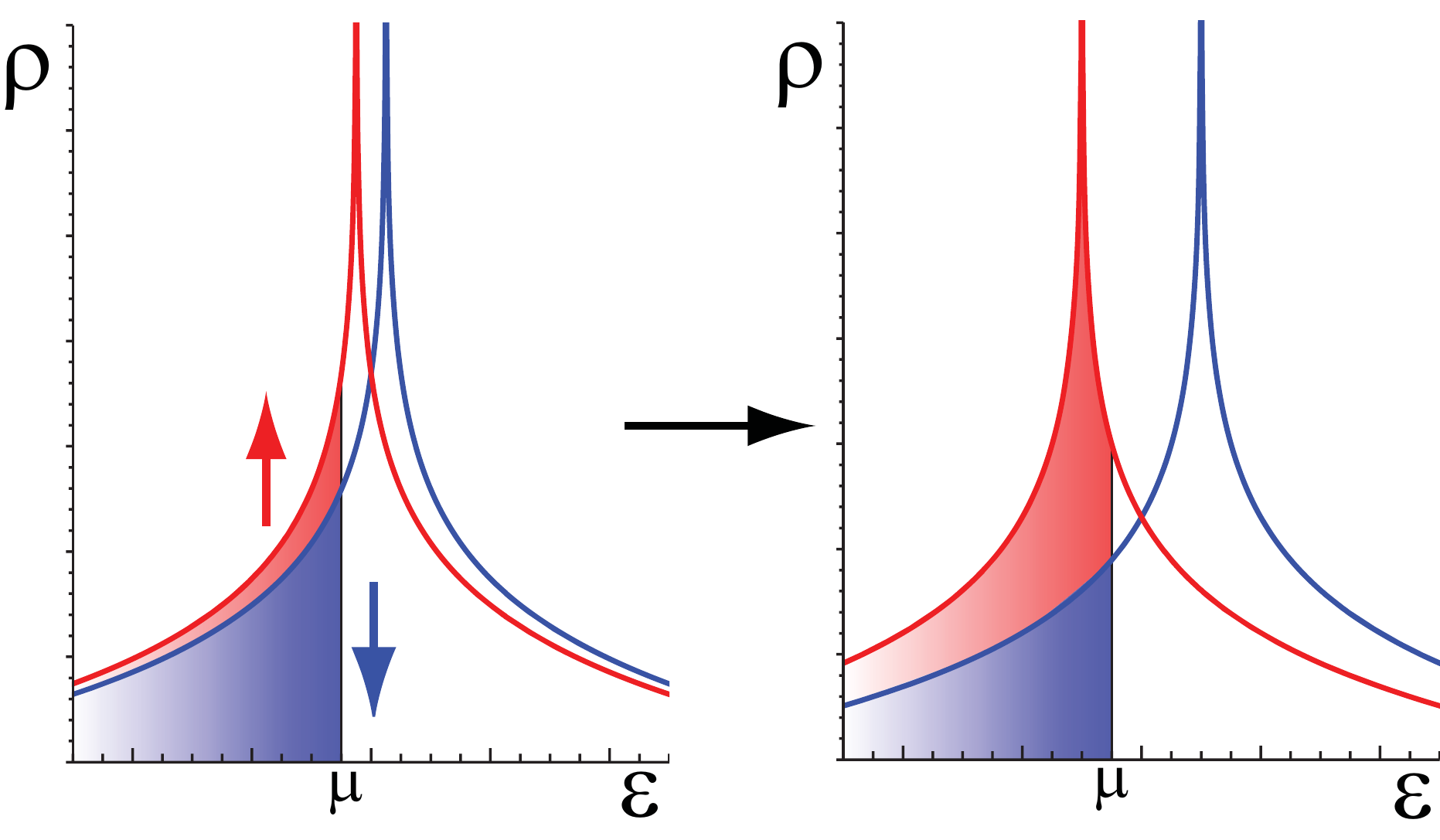}
\caption{\label{fig:cartoon} (Color online)  Cartoon for the formation of metamagnetism.  At the metamagnetic transition the majority band increases its filling through the van Hove singularity.  The presence of this singlarity leads to a reduced cost in single-particle energy allowing the gain in interaction energy to win out.}
\end{figure}

We now present a cartoon for the metamagnetic transition.  If the Fermi energy of a system is placed such that it is near a peak in the DoS, but the Stoner criterion is not yet satisfied then the system will be in the paramagnetic state.  If a magnetic field is applied then the Fermi-surfaces are split, one moving towards the peak and one away from it as shown in Fig.\ref{fig:cartoon}.  When one Fermi-surface reaches a region of high enough DoS then the interaction energy gain due to magnetising becomes greater than the single-particle energy cost in splitting the Fermi-surfaces further and the system spontaneously magnetizes.  If the Fermi-surface jumps discontinuously over the peak then there is a first order transition in the magnetisation.  This is known to occur if the curvature of the DoS is high enough such that it satisfies the Wohlfarth-Rhodes criterion~\cite{Wohlfarth} $3\left(\rho_{\rm F}'\right)^2<\rho_{\rm F}\rho_{\rm F}''$.  Thus the Stoner model with a peak in the DoS produces a metamagnetic wing.

\begin{figure}
\centering
\includegraphics[width=3in]{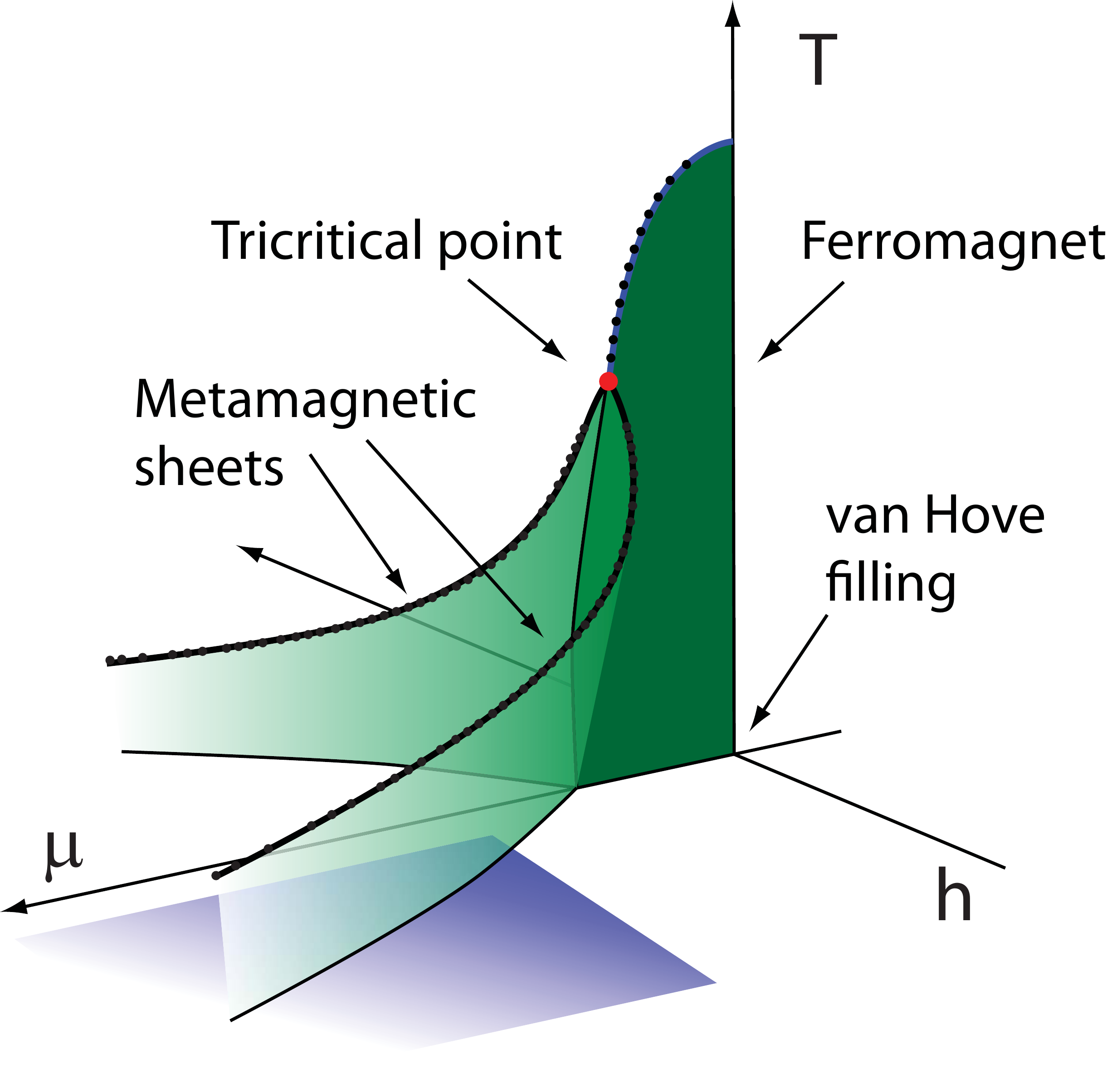}
\caption{\label{fig:nPD} (Color online)  The phase diagram for the Stoner model as a function of $\mu$, $h$ and $T$ for the next-nearest-neighbour tight-binding dispersion.  The line of second order transitions is given by $\partial^2_m F=0$, the line of metamagnetic critical endpoints by $\partial^3_m F=\partial^2_m F=0$ and the tricritical point by $\partial^4_m F=\partial^2_m F=0$.  The shaded area on the $\mu, h$ plane is a schematic representation of the region of this phase diagram which is observed in Sr$_3$Ru$_2$O$_7$.}
\end{figure}

The phase diagram for this model can be calculated from the mean-field free energy.  The phase transitions are determined by conditions on the derivatives of the free energy with respect to magnetization.  The second derivative gives the inverse of the magnetic susceptibility.  The susceptibility diverges at a second order transition, giving a condition for a continuous transition.  The line of metamagnetic critical endpoints is determined by the third derivative being zero.  By symmetry this derivative must be zero when there is no field and the tricritical point in the zero field plane is determined by the fourth derivative vanishing.  These derivatives are the coefficients of a Landau expansion of the free energy which we will derive shortly.  These conditions may be solved numerically to give the transition lines as a function of $\mu$, $h$ and $T$.  With the next-nearest-neighbour tight-binding dispersion these conditions give the phase diagram shown in Fig.\ref{fig:nPD}.  At zero field and close to van Hove filling there is a continuous transition into the ferromagnetic state as the temperature is decreased.  The transition becomes first-order at a tricritical point.  This first order transition opens up into two metamagnetic wings at non-zero field as $\mu$ gets further from van Hove filling.  These are sheets of discontinuous jumps in magnetization which end at a critical endpoint.

In Sr$_3$Ru$_2$O$_7$ the angle of the applied field tunes through the metamagnetic wing, effectively taking the role of $\mu$.  Rather than tuning the filling of a fixed band structure, the field angle controls the position of a peak in the DoS allowing tuning through the phase diagram by varying this angle.  We will consider how this may come about in a later section.  The parent tricritical point does not appear in the phase diagram of Sr$_3$Ru$_2$O$_7$ as a function of the angle of applied field~\cite{Yaguchi}.  Sr$_3$Ru$_2$O$_7$ can therefore be thought of as a segment of this phase diagram, schematically indicated in Fig. \ref{fig:nPD} (the experimental phase diagram is presented in Fig. \ref{fig:EPD}).

\subsection{Ginzburg-Landau expansion}

In order to calculate the phase diagram including spatial modulation we turn to a Ginzburg-Landau expansion of the free energy around the line of metamagnetic critical endpoints, an approach similar to that of Ref. \onlinecite{Millis,Green05}.  We begin by considering the general properties of Ginzburg-Landau expansions before developing an explicit expansion of the microscopic Hamiltonian.  We will see that the coefficients of the expansion are constrained by the fact that we are expanding about the line of critical endpoints.  The homogeneous phase diagram is calculated.  We will then investigate the formation of inhomogeneous transverse magnetisation. We will see that modulated phases become favourable along the line of critical endpoints.

\begin{figure}
\centering
\includegraphics[width=3in]{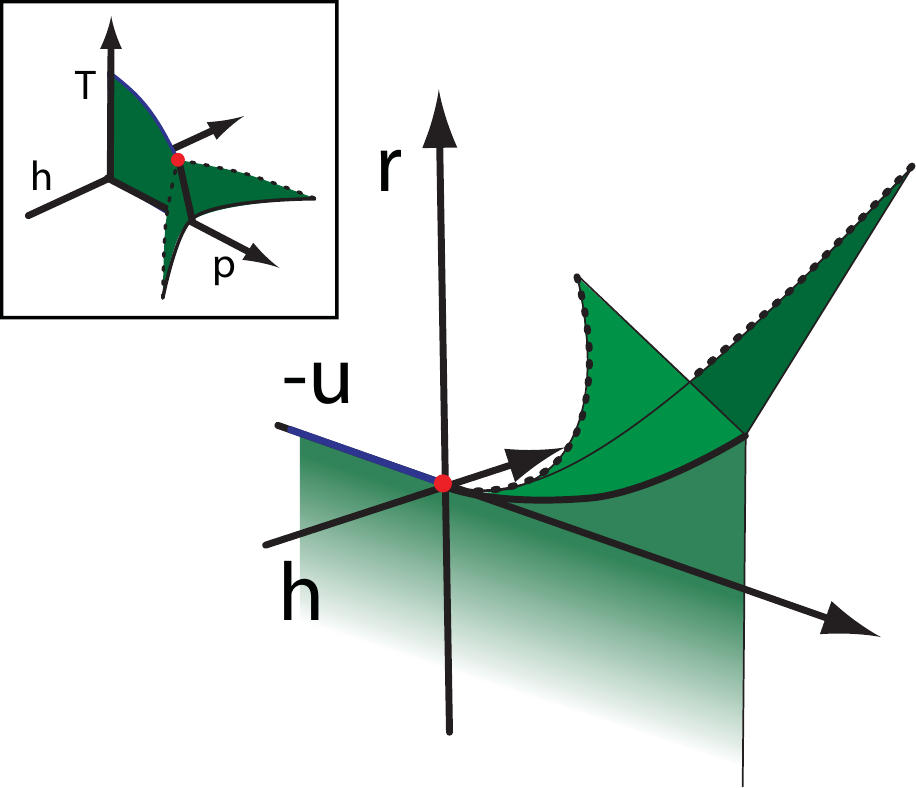}
\caption{\label{fig:LPD} (Color online)  The phase diagram for a general Landau expansion.  The inset shows how this phase diagram maps onto the microscopic parameters of a theory.  This should be compared with the calculated phase diagram for the the Stoner model, Fig. \ref{fig:nPD}.}
\end{figure}

The Ginzburg-Landau expansion is a method of studying phase transitions which relies on the smallness of the order parameter near to a continuous transition.  In this regime the free energy can be expanded in powers of the order parameter.  The terms which appear in this expansion are governed by the symmetry of the system.  In the case of the zero-field paramagnet to ferromagnet transition the free energy becomes an expansion in even powers of magnetisation
\begin{eqnarray}
\beta F_{\rm L}
&=&
r{\bf M}^2
+u{\bf M}^4
+v{\bf M}^6
-{\bf h}\cdot{\bf M}.
\label{L}
\end{eqnarray}
This gives a phase diagram in terms of the coefficients $r$, $u$ and the field ${\bf h}$, shown in Fig. \ref{fig:LPD}.  For $r<0$, $u>0$ the plane $h=0$ defines a sheet of first order transitions with the line $r=0$ being a second order transition line.  The point $r=u=h=0$ is a tricritical point where the second order transition bifurcates into two lines of critical endpoints determined by the conditions $r=\frac{9u^2}{20v}$, $h=\pm \frac{6u^2}{25}\sqrt{\frac{3|u|}{10v^3}}$. These transitions may be mapped onto the parameters of any given microscopic theory by calculating the coefficients $r$, $u$ and $v$ in terms of the microscopic parameters of the Hamiltonian.  For the case of the Stoner model we will calculate these coefficients as functions of $\mu$, $h$ and $T$.

In order to include the energy cost for spatial modulation we add terms involving gradients of the order parameter to the expansion.  These gradient terms become an expansion in powers of ${\bf q}$.  Like the powers of the order parameter these are constrained by the symmetries of the system.  If the system has inversion symmetry, like Sr$_3$Ru$_2$O$_7$, then modulations with wavevectors $\pm {\bf q}$ must have the same energy.  There can only be even powers of ${\bf q}$ in the expansion.  Inhomogeneity in this model is not the result of a Dzyaloshinskii-Moriya interaction~\cite{Moriya,Dzyaloshinskii,Bak}. Furthermore the direction of ${\bf q}$ will be picked out by anisotropies in the electronic dispersion.  We do not expect these anisotropies to alter the phase diagram and so will consider a simplified isotropic model.  Assuming a single ${\bf q}$ the Ginzburg-Landau expansion for the modulated system has the form:
\begin{eqnarray}
\beta F_{\rm L}
&=&
\left(r+K{\bf q}^2+L{\bf q}^4\right){\bf M}^2
+u_{\bf q}{\bf M}^4
+v{\bf M}^6
-{\bf h}\cdot{\bf M}.
\nonumber\\
\label{GL}
\end{eqnarray}

This is the same method often used to study the LOFF transition~\cite{Loff_review}.  We may therefore anticipate that a similar effect occurs in magnetization, where the transition between low and high magnetization states on the metamagnetic wing is split by the formation of an inhomogeneous magnetic phase.  In LOFF the expansion of the microscopic theory reveals that the $K$ coefficient of the expansion is approximately proportional to the $u$ coefficient.  Therefore the tricritical point, where $u=0$, is the point at which inhomogeneity appears.  We will look for a similar relationship in the case of Stoner magnetism.  It will emerge that in this case the relationship is more subtle.  The formation of modulated transverse magnetization self-consistently drives the longitudinal $M^4$ term negative.  This results in a reconstruction of the metamagnetic line to include a tricritical point accompanied by transverse spatial modulation.

We now turn to detailed calculation to obtain the Ginzburg-Landau expansion from the Hamiltonian Eq.(\ref{Ham}).  This process follows several standard steps.  The partition function is written as a path integral and the interaction terms decoupled.  The action is then expanded in powers of {\bf m} and {\bf q} to obtain the terms of the Ginzburg-Landau expansion.

The partition function is written as a path integral over Grassman fields,
\begin{eqnarray}
\mathcal{Z}
&=&
\int \mathcal{D} \left(\psi^\dagger, \psi\right)e^{-\int_0^\beta d\tau \left[\psi^\dagger \partial_\tau \psi+\hat{H}-\mu\hat{N}\right]}.
\end{eqnarray}
The interaction terms are decoupled by a Hubbard-Stratonovich transformation with coupling ${\bf m}({\bf x})\cdot\left(\psi^\dagger_{\bf x}{\boldsymbol \sigma}\psi_{\bf x}\right)$.  Integrating over the fermionic fields the partition function can be expressed as a field integral, $\mathcal{Z}=\int \mathcal{D}{\bf m} \; e^{-\mathcal{S}\left[{\bf m}\right]}$ where the Euclidean time action takes the form
\begin{eqnarray}
\mathcal{S}[{\bf m}]
&=&
\frac{g}{4} \int dy ~{\bf m}^2-{\rm tr} \ln{\left[\hat{G}_0^{-1}+\frac{g}{2}{\boldsymbol \sigma}\cdot{\bf m}\right]}
\nonumber\\
&=&
\frac{g}{4} \int dy ~{\bf m}^2-{\rm tr} \ln{\left[\hat{G}_0^{-1}\right]}
\nonumber\\
&&
+\sum_{n=1}^\infty \frac{(-1)^n}{n} \left(\frac{g}{2}\right)^2 {\rm tr} \left[\hat{G}_0 {\boldsymbol \sigma}\cdot{\bf m}\right]^n,
\label{action}
\end{eqnarray}
where $\int dy \equiv \int_0^\beta d\tau \int d^d x$, $\beta = \left(k_{\rm B}T\right)^{-1}$ and $d$ is the spatial dimension.  Here $\hat{G}_0^{-1}=-\partial_\tau-\xi_{\bf k}+h\sigma_z$, where $\xi_{\bf k}=\epsilon_{\bf k}-\mu$, denotes the inverse Green's function of the non-interacting electron.  In the second line we have introduced the expansion of the action in powers of ${\bf m}$.  It will be convenient to absorb the external field into a shift of the longitudinal magnetization, $m \mapsto m'=m+\frac{2}{g} h$.

The Landau theory is developed as an expansion around the saddle point of this action along the line of critical end-points. Varying the action with respect to ${\bf m}$ and applying the ansatz ${\bf m}=\bar M$, constant, gives the equation for the uniform saddle-point
\begin{eqnarray}
\bar M-\frac{2}{g} h
&=&
\frac{1}{\beta L^d} \sum_{k \sigma} \sigma G_\sigma(k)
\nonumber\\
&=&
\frac{1}{L^d}\sum_{{\bf k}\sigma} \sigma\, n_{\rm F}
\left[
\epsilon_{\bf k}-g \bar M\sigma/2
\right],
\label{cond1}
\end{eqnarray}
where $G_\sigma(k)=(i\omega_n-\xi_{\bf k}+\frac{g}{2} \bar M \sigma)^{-1}$ and $n_{\rm F}(\epsilon)=(e^{\beta(\epsilon-\mu)}+1)^{-1}$ is the Fermi distribution function with $\sigma=\pm 1$ denoting spin-up and -down electrons.  For a given interaction $g$, this gives the value of the saddle-point magnetization $\bar M$ as a function of chemical potential $\mu$, magnetic field $h$ and temperature $T$.

As mentioned earlier, the coordinates of the metamagnetic critical endpoint are found by the requirement that the second and third derivatives of the free energy with respect to magnetisation are zero. This gives the conditions
\begin{eqnarray}
&&\frac{2}{g}=-\frac{1}{L^d}\sum_{{\bf k}\sigma} 
n_{\rm F}^{(1)}\left[ \epsilon_{\bf k}-(g \bar M/2+h)\sigma \right],
\nonumber\\
&&
0=\frac{1}{L^d}\sum_{{\bf k}\sigma} \sigma 
n_{\rm F}^{(2)}\left[\epsilon_{\bf k}
-(g \bar M/2+h)\sigma \right],
\label{cond2}
\end{eqnarray}
where $n_{\rm F}^{(n)}(\epsilon)=\partial_{\epsilon}^n n_{\rm F}(\epsilon)$.  These equations not only determine the phase transitions but will be used to simplify the expressions for the coefficients of the Landau expansion.

\subsection{Homogeneous expansion}

The Ginzburg-Landau expansion is constructed by evaluation of the terms in the expansion of the action (\ref{action}).  As the expansion is centered on the line of critical endpoints we expand about the saddle-point value $\bar M$ and not zero.  The presence of the field singles out a particular direction in space and components of the magnetization parallel and perpendicular to this direction may have different properties.  We set ${\bf M} =(m+\bar M)\hat{\bf e}_\parallel+{\bf m}_\perp$ with the deviation from the saddle-point solution $\bar M$ presumed small.  Discarding the constant contribution to the action, the saddle-point solution ensures that most of the terms at first order in ${\bf M}$ must vanish, leaving only the field-dependent contribution,
\begin{equation}
{\mathcal S}^{(1)}
=
-\int d^dx \; h m.
\end{equation}
At second order the action can be split into longitudinal and transverse components,  
\begin{eqnarray}
{\mathcal S}^{(2)}
&=&
\frac{g}{4} \int d^dx \; {\bf m}^2
\nonumber\\
&&+\left(\frac{g}{2}\right)^2 {\rm tr} \left[ \hat G_\uparrow {\bf m}_\perp \cdot \hat G_\downarrow {\bf m}_\perp \right.
\nonumber\\
&&\left.+ \frac{1}{2}\left( (\hat G_\uparrow m)^2+(\hat G_\downarrow m)^2 \right) \right].
\end{eqnarray}
Defining the longitudinal and transverse susceptibilities
\begin{eqnarray}
\Pi_{|| \sigma}({\bf q})
&=&
\frac{1}{\beta L^d} \sum_k G_\sigma({\bf k}) G_\sigma({\bf k+q}),
\nonumber\\
\Pi_\perp ({\bf q})
&=&
\frac{1}{\beta L^d} \sum_k G_\uparrow({\bf k}) G_\downarrow({\bf k+q}),
\label{susceptibility}
\end{eqnarray}
we have
\begin{eqnarray}
{\mathcal S}^{(2)}
&=&
{\mathcal S}_{||}^{(2)}+{\mathcal S}_\perp^{(2)},
\nonumber\\
{\mathcal S}_{||}^{(2)}
&=&
\frac{g}{4}\int d^dx m^2 + \frac{g^2}{8} \sum_{q \sigma} \Pi_{|| \sigma} m_q m_{-q},
\nonumber\\
{\mathcal S}_{\perp}^{(2)}
&=&
\frac{g}{4}\int d^dx m_\perp^2 + \frac{g^2}{4} \sum_{q } \Pi_{\perp} {\bf m}_{\perp q} {\bf m}_{\perp -q}.
\end{eqnarray}
Expanding in powers of ${\bf q}$ will lead to the gradient terms in the Ginzburg-Landau expansion.  Initially we will consider the ${\bf q}=0$ homogeneous case.  Evaluating the susceptibilities we find
\begin{eqnarray}
{\mathcal S}_{||}^{(2)}
&=&
\beta \int d^dx \; r m^2,
\nonumber\\
{\mathcal S}_{\perp}^{(2)}
&=&
\beta \int d^dx \; r_\perp {\bf m}_\perp^2,
\end{eqnarray}
where
\begin{eqnarray}
r
&=&
\frac{g}{4}+\frac{g^2}{8} \frac{1}{L^d} \sum_{\bf k \sigma} n_{\rm F}^{(1)}(\epsilon_{\bf k}-g\bar M\sigma/2),
\nonumber\\
r_\perp
&=&
\frac{g}{4}-\frac{g^2}{4}\frac{1}{g \bar M}\frac{1}{L^d} \sum_{\bf k} \sigma n_{\rm F}(\epsilon_{\bf k}-g\bar M\sigma/2).
\end{eqnarray}
Considering the situation of zero field, $\bar M\rightarrow 0$.  $r_\perp$ becomes a constant and $r$ becomes $\frac{g}{4}\left( 1 + \frac{g}{L^d} \sum_{{\bf k}} n^{(1)}_{\rm F}(\epsilon_{\bf k})\right)$.  The condition $r=0$ corresponds to a second-order transition.  This is the standard Stoner criterion, $1=g \rho(\epsilon_{\rm F})$ at zero temperature.

At third order, the longitudinal and transverse magnetisations become coupled.  It is this coupling which allows inhomogeneity in the transverse component to affect the phase diagram of the longitudinal magnetisation.
\begin{eqnarray}
{\mathcal S}^{(3)}
&=&
-\frac{g^3}{24} {\rm tr} \left[ (\hat G_\uparrow m )^3 - (\hat G_\downarrow m)^3\right]
\nonumber\\
&&-\frac{g^3}{8} {\rm tr} \left[ \hat G_\uparrow^2 \hat G_\downarrow m {\bf m}_\perp^2 - \hat G_\uparrow \hat G_\downarrow^2 m {\bf m}_\perp^2 \right]
\nonumber\\
&=&
\beta \int d^dx \left[ s m^3 + s_\perp m {\bf m}_\perp^2 \right],
\end{eqnarray}
where
\begin{eqnarray}
s
&=&
-\frac{g^3}{48} \frac{1}{L^d} \sum_{{\bf k} \sigma} \sigma n_{\rm F}^{(2)}(\epsilon_{\bf k} - g \bar M\sigma/2),
\nonumber\\
s_\perp
&=&
-\frac{2}{(g \bar M)^2} \frac{1}{L^d} \sum_{{\bf k} \sigma} \left[ \sigma n_{\rm F}(\epsilon_{\bf k} - g \bar M \sigma/2)\right.
\nonumber\\
&& \;\;\;\;\;\;\;\;\;\;\;\;\;\;\;
\left.+ \frac{g \bar M}{2} n_{\rm F}^{(1)}(\epsilon_{\bf k} - g \bar M \sigma/2) \right].
\end{eqnarray}
$s$ vanishes in the absence of an external field, as expected from symmetry.  The line of critical endpoints is given by the condition $r=s=0$.

The expansion can be evaluated at each order to obtain the coefficients of the Landau expansion.  There is little to be gained by a detailed description of this process and we simply present the results here.  On the line of critical endpoints certain simplifications occur.  Applying the conditions (\ref{cond2}) to the expressions for the coefficients and adopting dimensionless magnetizations $\phi=m/\bar{M}$ and ${\boldsymbol \phi}_\perp={\bf M}_\perp/{\bar M}$, we obtain a Landau expansion with the form:
\begin{eqnarray}
\beta F_{\rm L}
&=&
h \bar M 
\big [
R\phi^2 + S\phi^3 + U\phi^4 + T\phi^5 + V\phi^6 -H \phi
\nonumber\\
&&
\;\;\;\;\;\;\;\;\;\;\;
+ R_\perp {\boldsymbol \phi}_\perp^2 + U_\perp {\boldsymbol \phi}_\perp^4 + V_\perp {\boldsymbol \phi}_\perp^6
\nonumber\\
&&
\;\;\;\;\;\;\;\;\;\;\;
+ S_{1} \phi {\boldsymbol \phi}_\perp^2 + U_{1} \phi^2 {\boldsymbol \phi}_\perp^2 
+ T_{1} \phi^3 {\boldsymbol \phi}_\perp^2  
\nonumber\\
&&
\;\;\;\;\;\;\;\;\;\;\;
+ V_{1} \phi^4 {\boldsymbol \phi}_\perp^2 + T_{2} \phi {\boldsymbol \phi}_\perp^4 
+ V_{2} \phi^2 {\boldsymbol \phi}_\perp^4 
\big]
\label{mmGL}
\end{eqnarray}
where
\begin{eqnarray}
& &
R=0,
\;\;\;\;\;\;
S=0,
\nonumber\\
& &
U
=
\frac{1}{4!}
\left(\frac{\bar M^3}{h} \right)
 \left( \frac{g}{2} \right)^4 
  \frac{1}{L^d}
\sum_{{\bf k} \sigma}
n_{\rm F}^{(3)}(\epsilon-g \bar M \sigma/2),
\nonumber\\
& &
T
=
\frac{1}{5!} 
\left(\frac{\bar M^4}{h} \right)
\left( \frac{g}{2} \right)^5 
 \frac{1}{L^d}
\sum_{{\bf k} \sigma}
\sigma n_{\rm F}^{(4)}(\epsilon-g \bar M \sigma/2),
\nonumber\\
& &
V
=
\frac{1}{6!}
\left(\frac{\bar M^5}{h} \right)
\left(\frac{g}{2}\right)^6
 \frac{1}{L^d}
\sum_{{\bf k} \sigma}
n_{\rm F}^{(5)}(\epsilon-g \bar M \sigma/2),
\nonumber\\
& &
R_{\perp}=\frac{1}{2},
\;\;\;\;\;\;
U_{\perp}
=
-\frac{1}{8},
\;\;\;\;\;\;
V_{\perp}
=
\frac{1}{16},
\nonumber\\
& &
S_{1}
=
-\frac{1}{2},
\;\;\;\;\;\;
U_{1}
=
-4 U_{\perp}
=
\frac{1}{2},
\nonumber\\
& &
T_{1}
=
-\frac{1}{2 }
+2U,
\;\;\;\;\;\;
T_{2}
=
\frac{3}{8},
\nonumber\\
& &
V_{1}
=
1-2U-\frac{5}{2}T,
\;\;\;\;\;\;
V_{2}
=
-\frac{3}{4}+\frac{3}{2}U.
\label{coeff}
\end{eqnarray}
The condition that we lie on the line of critical end-points is sufficient to reduce many of the coefficients of the expansion to constants.  This independence of the coefficients from the details of the dispersion and filling suggests that we may be able to deduce them from general principles.  We shall later show that this is the case.

These coefficients are sufficient to determine the phase diagram for homogeneous magnetisation as shown in Fig.\ref{fig:nPD}.  It is not favourable to form transverse magnetisation in the homogeneous case as the coefficients of the $\phi_\perp$ terms raise the free energy.  There is a metamagnetic transition in the longitudinal magnetisation, as there is when the transverse component is not considered.  We will show that it may become favourable to form transverse magnetisation when this magnetisation is spatially modulated and that the appearance of this component causes the reconstruction of the longitudinal metamagnetic transition.

\subsection{Transverse susceptibility}

Having discussed how features in the DoS can lead to metamagnetism we show that they can also lead to spatial modulation of the magnetism.

\begin{figure*}
\centering
\includegraphics[width=\textwidth]{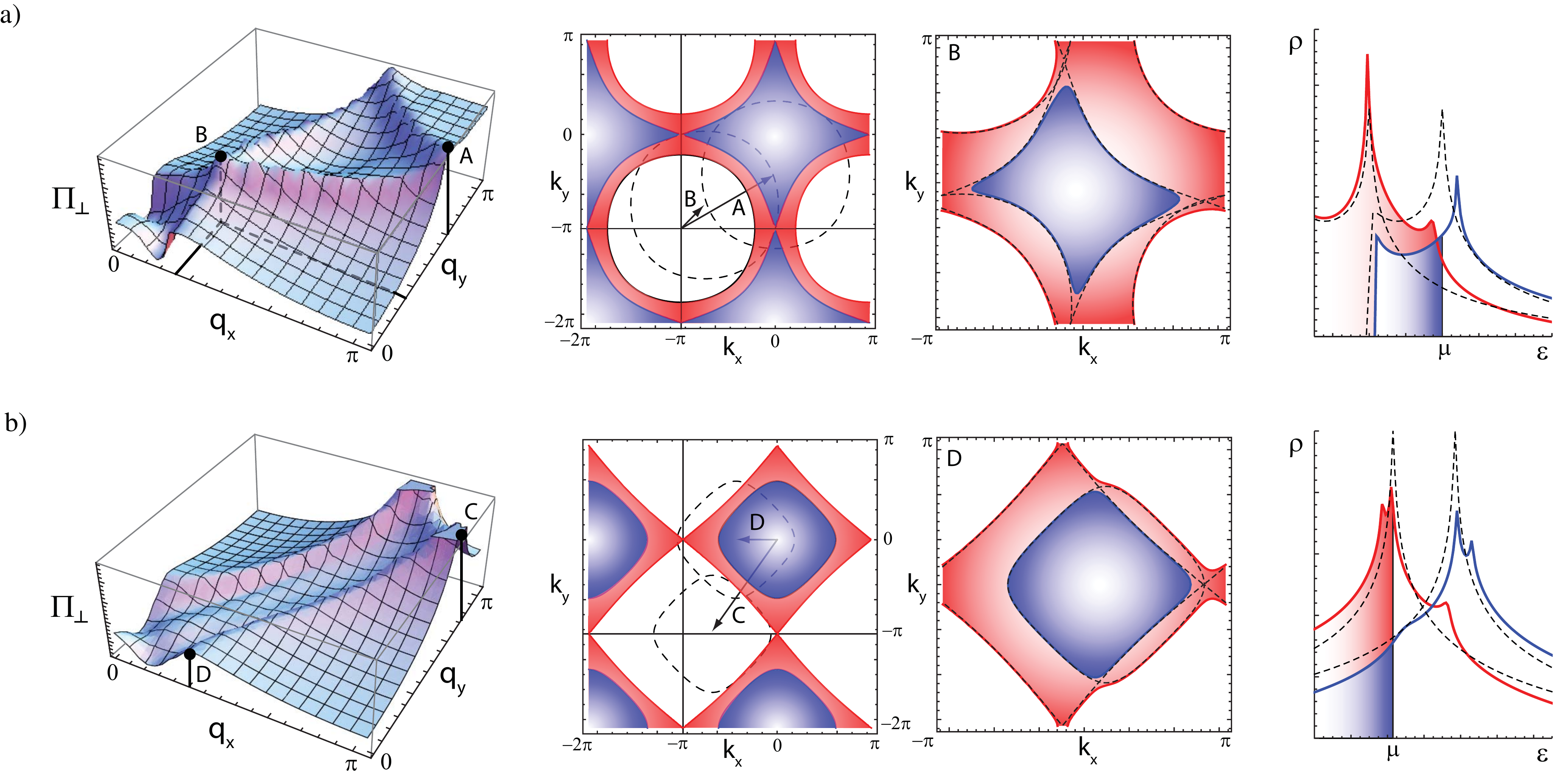}
\caption{\label{fig:susc} (Color online)  The transverse susceptibility and associated features in the Fermi surface and DoS of the next-nearest-neighbour tight-binding model discussed in the text.  The first column shows transverse susceptibility, $\Pi_{\perp}({\bf q})$ evaluated for two situations, a) $t=0.8$, $\mu=-0.6$, and b) $t=0.2$, $\mu=-0.37$, chosen to illustrate various possible wavevectors of disortion.  In both cases $\bar M$ is chosen such that the spin-splitting places one species' Fermi surface at the van Hove singularity, in a) this is the minority Fermi surface and in b) the majority.  The susceptibility shows peaks at both large and small ${\bf q}$.  In the second column the spin-up and -down Fermi surfaces are shown for each case. Displacements of these Fermi surfaces corresponding to the wavevectors of the peaks in the susceptibility are indicated.  These displacements show the origin of the peaks.  Wavevectors A, B and C are all nesting vectors, however D is not, it corresponds to a different situation described below.  Note that in a) we have indicated the equivalent displacement of the hole rather than electron pocket for ease of visualisation.  The third column shows the Fermi surfaces after hybridisation for the low {\bf q} cases, these are the Fermi surfaces in the spiral state.  The wavevector D corresponds to opening the neck of the Fermi surface across the van Hove singularity.  The black dotted line shows the Fermi surface before hybridization.  The rightmost column shows the densities of states for the two spin-species before (black dotted line) and after (solid lines) the spiral state forms.  In a) a new peak has appeared below the Fermi level.  In b) the peaks in the densities of states have split and moved below the Fermi level.  The energetic origin of the nesting (A, B, C) and non-nesting (D) vectors is therefore similar, being associated with the appearance of new peaks in the DoS below the Fermi surface.}
\end{figure*}

The tendency of the system to form order at any wavevector is encoded in the ${\bf q}$-dependent susceptibility.  We will consider modulation in the component of the magnetisation transverse to the applied field.  While spatial modulation of the longitudinal magnetization is possible, it does not lead to the type of phase reconstruction considered here and we do not study it.  The susceptibility for transverse magnetisation is given by Eq.(\ref{susceptibility}).  Putting explicitly the ${\bf q}$-dependence, which we previously ignored, this becomes:
\begin{eqnarray}
\Pi_\perp({\bf q})
&=&
\sum_{\bf k} \frac{n_{\rm F}\left(\epsilon_{{\bf k}+{\bf q}}-\frac{g\bar{M}}{2}\right)-n_{\rm F}\left(\epsilon_{\bf k}+\frac{g\bar{M}}{2}\right)}{\epsilon_{\bf k}-\epsilon_{{\bf k}+{\bf q}}+g\bar{M}}
\nonumber\\
\label{susc}
\end{eqnarray}
where $n_{\rm F}\left(\epsilon\right)=\left(1+e^{\beta\left(\epsilon-\mu\right)}\right)^{-1}$ is the Fermi-Dirac distribution, ${\bf q}$ is the wavevector of modulation and $\bar{M}$ is a shifted magnetisation which includes the external field: $\bar{M}\mapsto\bar{M}'=\bar{M}+\frac{2}{g}h$.  This may be numerically evaluated for all ${\bf q}$ at any point in the phase diagram ($\mu$, $\bar{M}$, $T$).  The first column of Fig.\ref{fig:susc} shows the susceptibility evaluated for two points near the metamagnetic transition of our example tight-binding dispersion.  Parameters are chosen to illustrate the various situations which may arise.  These plots show that the susceptibility has peaks both at high and low wavevector.  A similar investigation has been presented by Monthoux and Lonzarich~\cite{Monthoux}.  The subsequent calculations will be performed with the parameters of Fig. \ref{fig:susc}b.

\begin{figure*}
\centering
\includegraphics{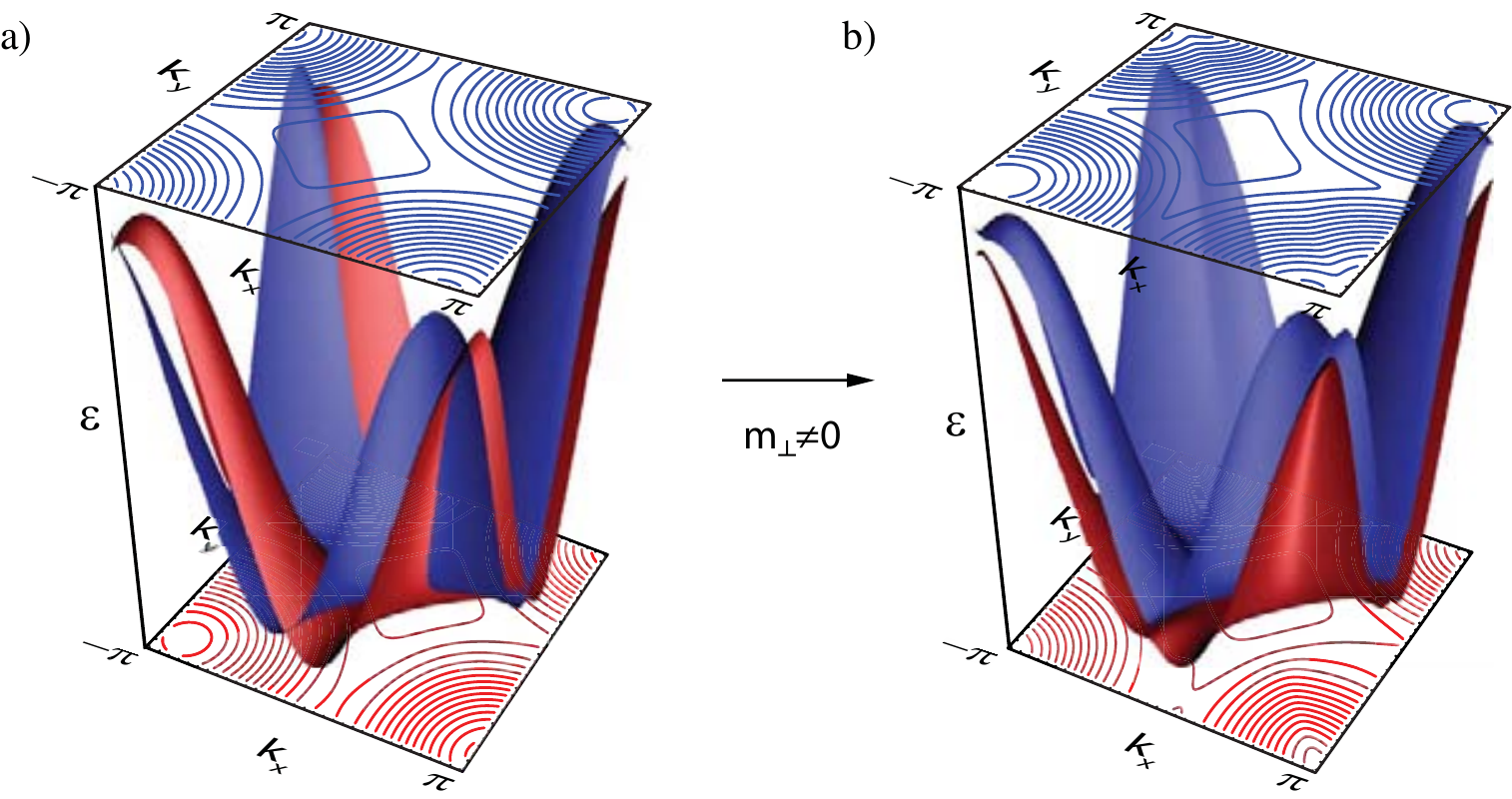}
\caption{\label{fig:anticrossing} (Color online)  a) Energy dispersions for the spin-up (red [dark grey, lower plot]) and spin-down (blue [light grey, upper plot]) Fermi surfaces shifted by $\pm {\bf q}/2$.  b) Energy dispersions including a transverse magnetization.  This hybridises the bands, causing anticrossing and the formation of new saddle points.}
\end{figure*}

In order to get a physical picture of why inhomogeneous magnetisation is favourable at particular wavevectors we will consider the way in which the energy dispersions are altered by inhomogeneity and how this appears in terms of peaks in the DoS.  The formation of a spiral magnetisation state involves hybridizing spin-up and -down electrons.  The dispersion of the hybridised state is given by
\begin{eqnarray}
2 E^\pm_{\bf k}
&=&
\left(\epsilon_{{\bf k}+{\bf q}/2}+\epsilon_{{\bf k}-{\bf q}/2}\right)
\nonumber\\
&&
\pm \sqrt{\left(\epsilon_{{\bf k}+{\bf q}/2}-\epsilon_{{\bf k}-{\bf q}/2}+g m_{||}+h\right)^2+\left(g m_\perp\right)^2},
\nonumber\\
\end{eqnarray}
where $m_{||}$ and $m_\perp$ are the components of magnetisation parallel and perpendicular to the applied field.  The second column of Fig.\ref{fig:susc} shows the undistorted Fermi surfaces shifted by $\pm {\bf q}$.  The development of transverse magnetisation then hybridises the bands.  The anti-crossing where the undistorted bands are degenerate results in the appearance of additional saddle-points in the dispersion, as is illustrated in Fig.\ref{fig:anticrossing}.  These saddle points produce additional peaks in the DoS as seen in Fig.\ref{fig:susc}.  If the new peak lies at or below the Fermi surface then it can result in a lowering of the energy.  By occupying the states under this new peak the formation of spiral magnetization reduces the energy cost of forming the transverse magnetization, allowing the interaction energy gain to win out.  We see from Eq.(\ref{susc}) that when a displacement of the Fermi surface by ${\bf q}$ overlaps significantly with the original Fermi surface then the denominator is minimised and the distortion is favoured.  This is due to large regions of the Fermi surface being removed by anticrossing of the dispersions, resulting in a peak in the DoS at the Fermi surface.

The various peaks in the susceptibility have different origins in the Fermi surface.  The high ${\bf q}$ peaks are related to the standard partial nesting wavevectors of the Fermi surface.  The small-${\bf q}$ peak in Fig. \ref{fig:susc}a corresponds to a new nesting vector between the spin-split Fermi surfaces.  This corresponds to nesting one Fermi surface inside the other and is illustrated in the center panel of Fig. \ref{fig:susc}a.  The low-${\bf q}$ maxima on the $q_x$ and $q_y$ axes of Fig. \ref{fig:susc}b do not correspond to nesting vectors.  They correspond to distortions which cause the Fermi surface of the spiral state to jump over one of the van Hove singularities of the original dispersion, as shown in the figure.  In all cases the distortions result in new peaks in the DoS.

We note that the ${\bf q}$ in this picture may be a substantial fraction of the Brillouin zone and depends on the details of the dispersion.

\subsection{Inhomogeneous expansion}

We have found the phase diagram for homogeneous magnetization from the expansion of the free energy and examined the possibility of inhomogeneous states from the susceptibility.  We continue in the spirit of a Ginzburg-Landau theory and expand the coefficients of the Landau expansion in powers of ${\bf q}$.  This approach is valid when considering small wavevector features.  As we will later argue it is not necessary for the structure of the phase diagram that the inhomogeneity is at small ${\bf q}$.  However, we will restrict our explicit analysis to cases where this expansion is valid and carry our conclusions across to the more general case with appropriate modifications.

In order to simplify our analysis, we study the simplest form of inhomogeneity, namely a spiral ${\boldsymbol \phi}_\perp({\bf r})=\phi_\perp \left(\cos{{\bf q}\cdot{\bf r}}, \sin{{\bf q}\cdot{\bf r}}\right)$, where ${\bf q}$ is the wavevector of the spiral. This is the magnetic analogue of the Fulde-Ferrell ansatz. As in LOFF the most stable  inhomogeneous phase may consist of spirals with several different ${\bf q}$-vectors superposed, creating a `crystalline' structure.  The competition between such states is complex and will not be considered here, but the ansatz is enough to identify the regions of inhomogeneity.  With this simplifying ansatz, the free energy has the form
\begin{widetext}
\begin{eqnarray}
\beta F_{\rm L}
&=&
h \bar M \left[
\left( R \phi^2 + U \phi^4 + T \phi^5 + V \phi^6 - H \phi \right)
\right.
\nonumber\\
&& 
\;\;\;\;\;\;\;
+
\left(
R_\perp +K_\perp {\bf q}^2 + L_\perp {\bf q}^4 +(S_1 + K_{1} {\bf q}^2) \phi + (U_1 + K_{2} {\bf q}^2)\phi^2 
+ T_1 \phi^3 + V_1 \phi^4 
\right) \phi_\perp^2
\nonumber\\
&&
\;\;\;\;\;\;\;
\left.
+
\left(
U_\perp + K_{3} {\bf q}^2 + T_2 \phi + V_2 \phi^2
\right) \phi_\perp^4
+ V_\perp \phi_\perp^6
\right],
\label{GL2}
\end{eqnarray}
\end{widetext}
where the terms $L_\perp$, $K_1$, $K_2$ and $K_3$ are required to bound the free energy for the case of non-zero {\bf q}.  In order to determine the gradient terms in the Ginzburg-Landau expansion, we return to the transverse susceptibility Eq.(\ref{susceptibility}) and allow for the previously-neglected {\bf q}-dependence. Expanding the Green's function gives
\begin{eqnarray}
G_\sigma({\bf k}+{\bf q})
=
G_\sigma({\bf k}) + [G_\sigma({\bf k})]^2 \partial_{k_i}\epsilon_{\bf k} q_i 
\nonumber\\
+ \left( [G_\sigma({\bf k})]^3 \partial_{k_i}\epsilon_{\bf k} \partial_{k_j}\epsilon_{\bf k} + \frac{1}{2} [G_\sigma({\bf k})]^2 \partial_{k_i, k_j}^2 \epsilon_{\bf k}\right)q_iq_j
\nonumber\\
+ O(q^3).
\end{eqnarray}
The first term in this expansion gives the homogeneous term already considered.  Terms of first order in ${\bf q}$ cannot contribute to the action due to the symmetry of the electronic dispersion.  Including second order terms we find the action
\begin{eqnarray}
{\mathcal S}_\perp^{(2)}
=
\beta \int d^dx \left[ r_\perp {\bf m}_\perp^2 + K_\perp (\nabla {\bf m}_\perp)^2 \right],
\end{eqnarray}
where $r_\perp$ is as before and
\begin{eqnarray}
K_\perp
=
-\frac{1}{4g \bar M^3} \frac{1}{L^d} \sum_{{\bf k} \sigma} \left[ \sigma n_{\rm F}(\epsilon_{\bf k}-g \bar M\sigma/2) \right.
\nonumber\\
\left.+ \frac{g\bar M}{2}n_{\rm F}^{(1)}(\epsilon_{\bf k}-g \bar M\sigma/2) \right] (\partial_{\bf k}\epsilon_{\bf k})^2.
\label{k}
\end{eqnarray}
Here we have used the fact that the angular dependence of the integrand enters only through the derivatives of the energy.  Assuming a dispersion with square symmetry allows us to use the relationship
\begin{eqnarray}
\int d^{d-1}k \; \partial_{k_i, k_j}^2 \epsilon_{\bf k} = \int d^{d-1}k \; \partial_{\bf k}^2 \epsilon_{\bf k} \frac{\delta_{ij}}{d}.
\end{eqnarray}
Although the coefficient $K_\perp$ is independent of the direction of {\bf q} this is not true for all coefficients.  Higher order terms such as $L_\perp$ (which is 4th order in ${\bf q}$) are anisotropic and pin the wavevectors to specific directions in the lattice.  Since we are considering a lattice with square symmetry, there will be at least four degenerate directions along which {\bf q}-vectors could lie.  The $K_1$, $K_2$ and $K_3$ terms may also be calculated by gradient expansions of the appropriate higher-order terms.  Here we neglect this complication.

\begin{figure}
\centering
\includegraphics[width=3in]{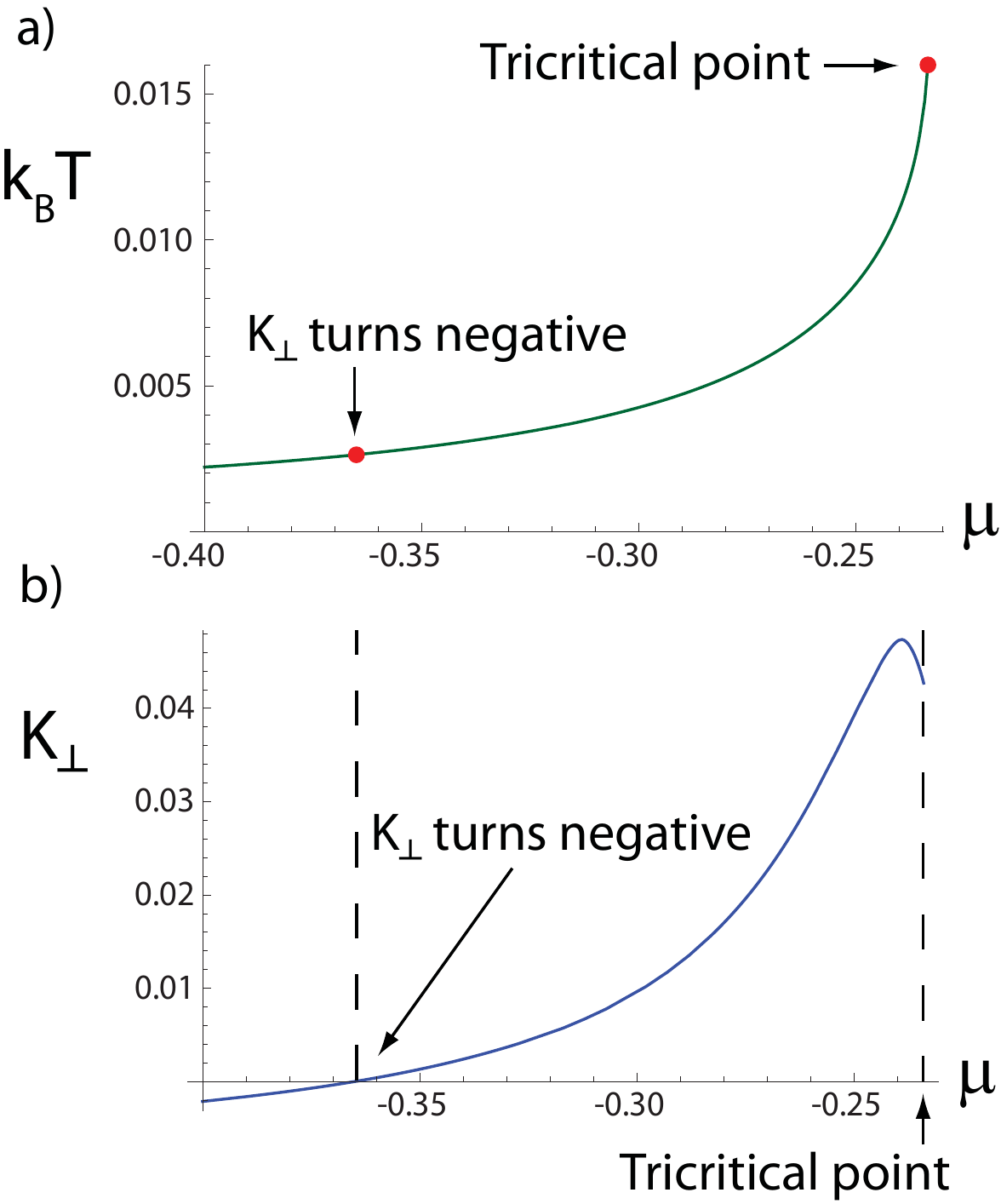}
\caption{\label{fig:kneg} (Color online)  a) The line of metamagnetic critical endpoints projected onto the $\mu$, $T$ plane.  b) $K_\perp$ evaluated along the line of metamagnetic critical endpoints.  As $\mu$ gets further from the van Hove singularity this becomes negative.  Plots evaluated for the dispersion discussed in the text with $t=0.2$ and $g=1.7$.}
\end{figure}

$K_\perp$ is the leading order tendency to formation of spiral magnetization.  We may evaluate this coefficient numerically at every point along the line of metamagnetic critical endpoints.  We find that for the example tight-binding dispersion it varies smoothly as we move along the transition and in fact becomes negative as we move further from the van Hove singularity, as shown in Fig.(\ref{fig:kneg}).  This shows an instability to the formation of spatially modulated magnetization.   We will use this fact to motivate the construction of a phase diagram for the formation of inhomogeneous magnetization along the metamagnetic transition.

We now consider how this relates to the analysis in the LOFF case.  If we linearise the Fermi velocity about the Fermi surface then the expansion of the longitudinal magnetization to quadratic order gives a coefficient $K \propto U$.  Since $U$ becomes negative at the tricritical point this indicates that modulated states become favourable at the tricritical point.  Rather than simply turning first-order the transition occurs via an inhomogeneous phase.  In this limit the equations for spatially modulated magnetization and superconductivity become equivalent~\cite{Rice,Loff_review}.  In the case which we are presently considering the metamagnetic transition is caused by proximity to van Hove singularities.  In this situation the dispersion cannot be linearised and the full form must be used instead.  In this case there is no simple relation between coefficients as in LOFF.

\section{Phenomenology}

\subsection{Phenomenological derivation of Ginzburg-Landau expansion}

\subsubsection{Metamagnetism}

Thus far we have given a heuristic description of the formation of inhomogeneous phases and derived a Ginzburg-Landau expansion for the formation of such states from an expansion of the Hamiltonian.  We now discuss how such a Ginzburg-Landau theory may be deduced on general grounds of symmetry and the topology of the metamagnetic phase diagram.  We will find that such arguments reproduce the form of the expansion and its coefficients as derived from microscopics.  We then calculate a phase diagram for such a theory.

The generic phase diagram of a metamagnetic system may be produced from a simple Landau expansion about zero magnetization, Eq.(\ref{L}).  We wish to investigate the potential for inhomogeneous order leading to a reconstruction of the metamagnetic wings and so develop an expansion about the line of metamagnetic critical end-points.  We denote the longitudinal magnetization in the vicinity of the line of metamagnetic critical end-points by $M=m+ \bar{M}$, where $\bar{M}$ is the magnetization on the line and $m$ is the deviation from it. The perpendicular component is denoted ${\bf M}_{\perp}$.  Substituting this shift of order parameter into the Landau free energy (\ref{L}) results in a theory with terms up to sixth order in $m$, $\bar{M}$ and ${\bf M}_\perp$.  This unwieldy number of terms is reduced considerably by explicitly constraining our expansion to be about the line of critical end-points.  The conditions that $\bar{M}$ be the magnetization along the line of critical
end-points are that the first three derivatives of the Landau free energy (\ref{L}) with respect to ${\bf M}$ are zero.  This gives 
\begin{eqnarray}
\partial_{\bf M} \left(\beta F_{\rm L}\right)
&=&
0
=
2r{\bf M} + 4u{\bf M}^3+6v{\bf M}^5-{\bf h},
\nonumber\\
\partial_{\bf M}^2 \left(\beta F_{\rm L}\right)
&=&
0
=
2r + 12u{\bf M}^2+30v{\bf M}^4,
\nonumber\\
\partial_{\bf M}^3 \left(\beta F_{\rm L}\right)
&=&
0
=
24u{\bf M}+120v{\bf M}^3,
\label{constraints}
\end{eqnarray}
which imply the relationships $r=15v \bar M^4$, $u=-5v \bar M^2$ and $h=16v \bar M^5$ when on the line of critical endpoints.  

We adopt dimensionless magnetizations as before.  Substituting into Eq.(\ref{L}) with the coefficients of the expansion constrained to lie along the line of critical end-points by Eq.(\ref{constraints}), we find the free energy (\ref{mmGL}) with
\begin{eqnarray}
S& = &0,
\;\;
U = 5/8,
\;\;
T = 3/8,
\;\;
V = 1/16,
\nonumber\\
R_\perp &=& 1/2,
\;\;
U_\perp = -1/8,
\;\;
V_\perp = 1/16,
\nonumber\\
S_{1} &=& -1/2,
\;\;
U_{1} = 1/2,
\;\;
T_{1} = 3/4,
\nonumber\\
V_{1} &=& 3/16,
\;\;
T_{2} = 3/8,
\;\;
V_{2} = 3/16.
\label{coeffs}
\end{eqnarray}
The values for the transverse coefficients, and most of the coupling coefficients, are identical to that found from the microscopic analysis Eq.(\ref{coeff}).  The longitudinal coefficients depend on the microscopics and vary along the line of critical endpoints, so would not be expected to be reproduced by this analysis.

Exactly on the line of critical end points, $R=0$ and $H=0$. We allow non-zero values in order to parametrize deviations from the line of critical end-points.  This Landau theory leads to a line of first order transitions at $R<0$ terminating at a second order end-point at $H=0$, $R=0$ corresponding to the metamagnetic wing of  Fig.~\ref{fig:LPD}. 

\subsubsection{Inhomogeneous magnetism}

We now wish to consider whether inhomogeneous magnetic states are more favourable over any region of the phase diagram than the homogeneous ferromagnetic or paramagnetic states.  Such inhomogeneous states will produce extra contributions to the free energy dependent on the magnetization gradient.  As we did previously, we consider only inhomogeneities in the transverse component of the magnetization.  Also, we restrict our study to cases where symmetry forbids terms linear in the gradient.

The first term that must be added to the free energy density is $K_\perp \left(\nabla {\boldsymbol \phi}_\perp({\bf r})\right)^2$.  When $K_\perp < 0$ higher order gradient terms are required in the free energy density: $L(\nabla^2 {\boldsymbol \phi}_\perp)^2$, $K_{2} \phi^2 \left(\nabla {\boldsymbol \phi}_\perp\right)^2$ and $K_{3} \left(\nabla {\boldsymbol \phi}_\perp\right)^2 {\boldsymbol \phi}_\perp$.

With the spiral ansatz ${\boldsymbol \phi}_\perp({\bf r})=\phi_\perp (\cos{{\bf q}\cdot{\bf r}}, \sin{{\bf q}\cdot{\bf r}})$ the free energy including the effects of spatial modulation reduces to Eq.(\ref{GL2}) with the homogeneous coefficients given by Eq.(\ref{coeffs}).

As we have shown from our microscopic calculation of $K_\perp$ for the tight-binding dispersion this coefficient varies smoothly along the line of metamagnetic critical endpoints and can eventually turn negative.  We will therefore use $K_\perp$ as a parameter which represents movement along the metamagnetic wing.   The formation of inhomogeneity will reconstruct the metamagnetic transition, as we will shortly show.  We will determine the phase diagram of the Ginzburg-Landau free energy as a function of $R$, $H$ and $K_\perp$. These parametrize directions within the metamagnetic wing, perpendicular to it and along the line of critical end points, respectively. Finally, we note that such a model has a fundamental anisotropy due to the influence of the lattice and, strictly, this is reflected in the higher order gradient terms in the free energy. This anisotropy will determine the direction of {\bf q}, but does not affect the topology of the phase diagram and we do not treat it explicitly.

\subsection{Phase diagram}

Having discussed how the Ginzburg-Landau expansion can be obtained we now calculate the phase diagram which this expansion gives.

\begin{figure}
\centerline{\includegraphics[width=3in]{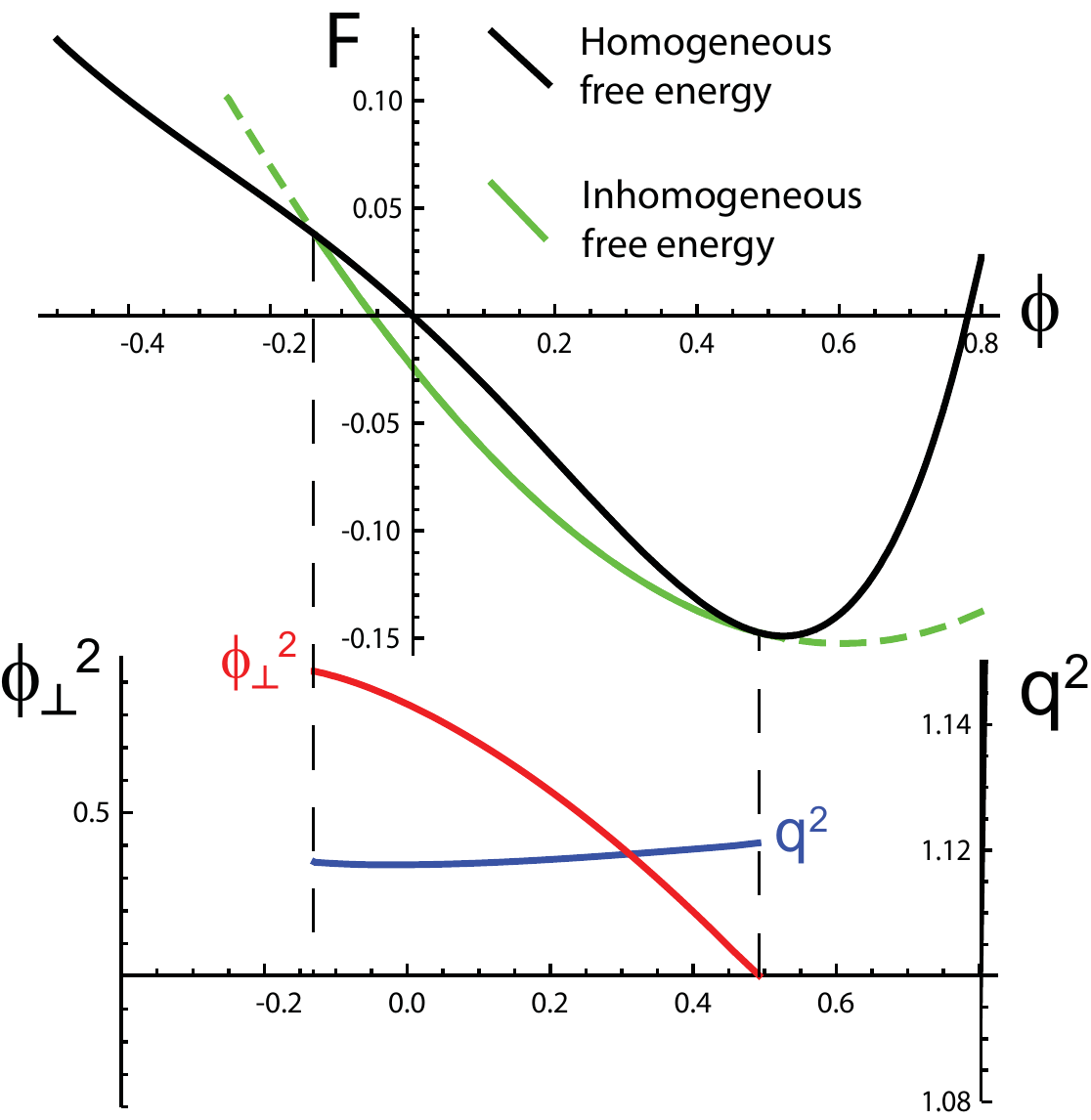}}
\caption{\label{fig:qfe} (Color online) Free energy with ${\bf q}$-dependent region.  Shown in black is the free energy when the magnetization is homogeneous.  The green (grey) line is the free energy when the magnetization is modulated.  The dotted regions show where this is unphysical, either increasing the free energy or corresponding to an imaginary magnetization.  In the solid region the magnetization is real and lowers the free energy.  Also shown are the transverse magnetization and wavevector in this region.} 
\end{figure}

Determining the phase diagram for Eq.(\ref{GL2}) involves locating the minima of the free energy as a function of $\phi$, $\phi_\perp$ and ${\bf q}$.  The broad scheme is as follows: Minimizing the free energy with respect to ${\bf q}$ gives the optimum wavevector ${\bf \bar q}(\phi, \phi_\perp)$.
Focusing on this wavevector, minimization of the free energy with respect to $\phi_\perp$ gives the optimum inhomogeneous transverse magnetization ${\bf \bar \phi_\perp} (\phi)$.  There is no real solution for ${\bf \bar \phi_\perp} (\phi)$ over much of the phase diagram.  This leads to a restricted region where inhomogeneity is allowed.  This corresponds to the region in which the inhomogeneous terms of the free energy lower the total free energy, rather than raise it.  An example plot of the free energy is shown in Fig.\ref{fig:qfe}.  The value of $\phi$ which minimizes the free energy therefore determines the longitudinal and transverse magnetization, as well as the wavevector of the inhomogeneous magnetization.   

There are many terms in the free energy and it is hard at first glance to understand the role of the various terms and how the phase diagram comes about.  To clarify this we have performed an extensive study of the role of the various terms which we present in the appendix.

\begin{figure}
\centering
\includegraphics[width=3in]{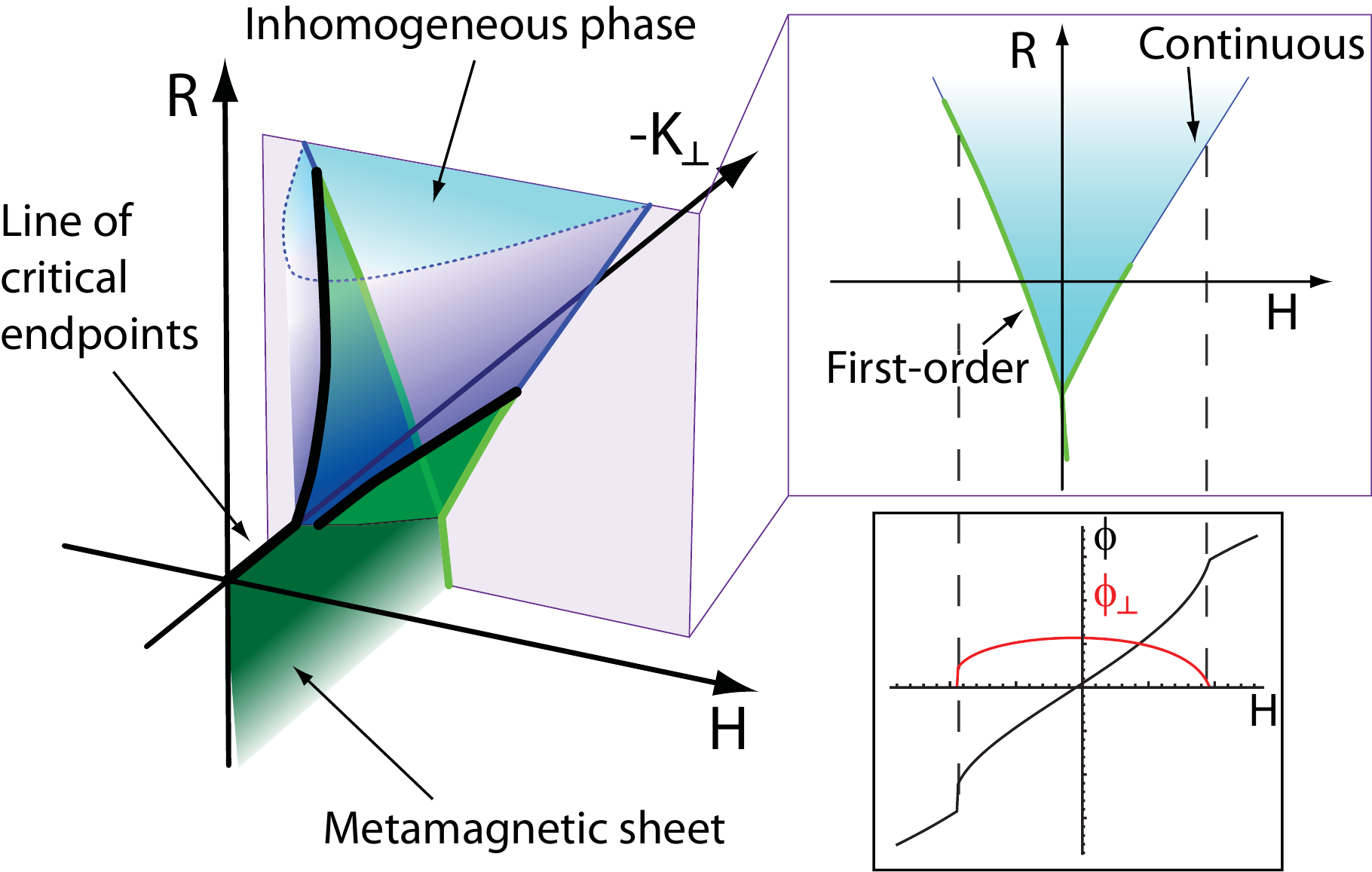}
\caption{\label{fig:PD} (Color online)  Phase diagram for the Ginzburg-Landau theory.  Green (dark grey) sheets represent first-order transitions in $\phi$.  Blue (pale grey) sheets represent continuous transitions into the inhomogeneous phase.  A cut through the phase diagram at constant $K_\perp$ and the variation of $\phi$ and $\phi_\perp$ along a path through this cut are shown.  This shows both first-order and continuous transitions.}
\end{figure}

Carrying out the minimization analysis results in the phase diagram indicated in Fig.\ref{fig:PD}. In this figure, the line $R=H=0$ is the parent line of metamagnetic critical end points. Upon moving along this line away from the tricritical point, $K_\perp$ reduces from a positive value, eventually becoming negative. When it becomes sufficiently negative, the metamagnetic sheet bifurcates into two wings. This structure is symmetry broken, or dislocated, in that the smaller first-order wing at higher fields does not emerge from the metamagnetic sheet at the same point as the larger wing.  This phase diagram is very similar to the dislocated tricritical point of Green {\it et al}~\cite{Green05}.  Here the bifurcation is caused by the appearance of the inhomogeneous phase which provides a `roof' to the anomalous phase as a continuous transition into the phase, which is missing in the previous theory.  Transitions into this inhomogeneous region occur in two different ways: In the first case, indicated by the green (dark grey) wings in Fig.\ref{fig:PD}, the transition is first order in both the longitudinal and transverse magnetization with a step change in the former and the latter appearing discontinuously from zero (and at finite ${\bf q}$). The second type of transition into the inhomogeneous region is indicated by the blue (pale grey) sheet in Fig.\ref{fig:PD}. On this sheet the transverse magnetization undergoes a second order transition, appearing continuously from zero. The longitudinal magnetization undergoes a continuous transition with a step change in its gradient upon moving through this sheet.  This kink is the `ghost' of the transition in the transverse magnetization.

\begin{figure}
\centering
\includegraphics[width=3in]{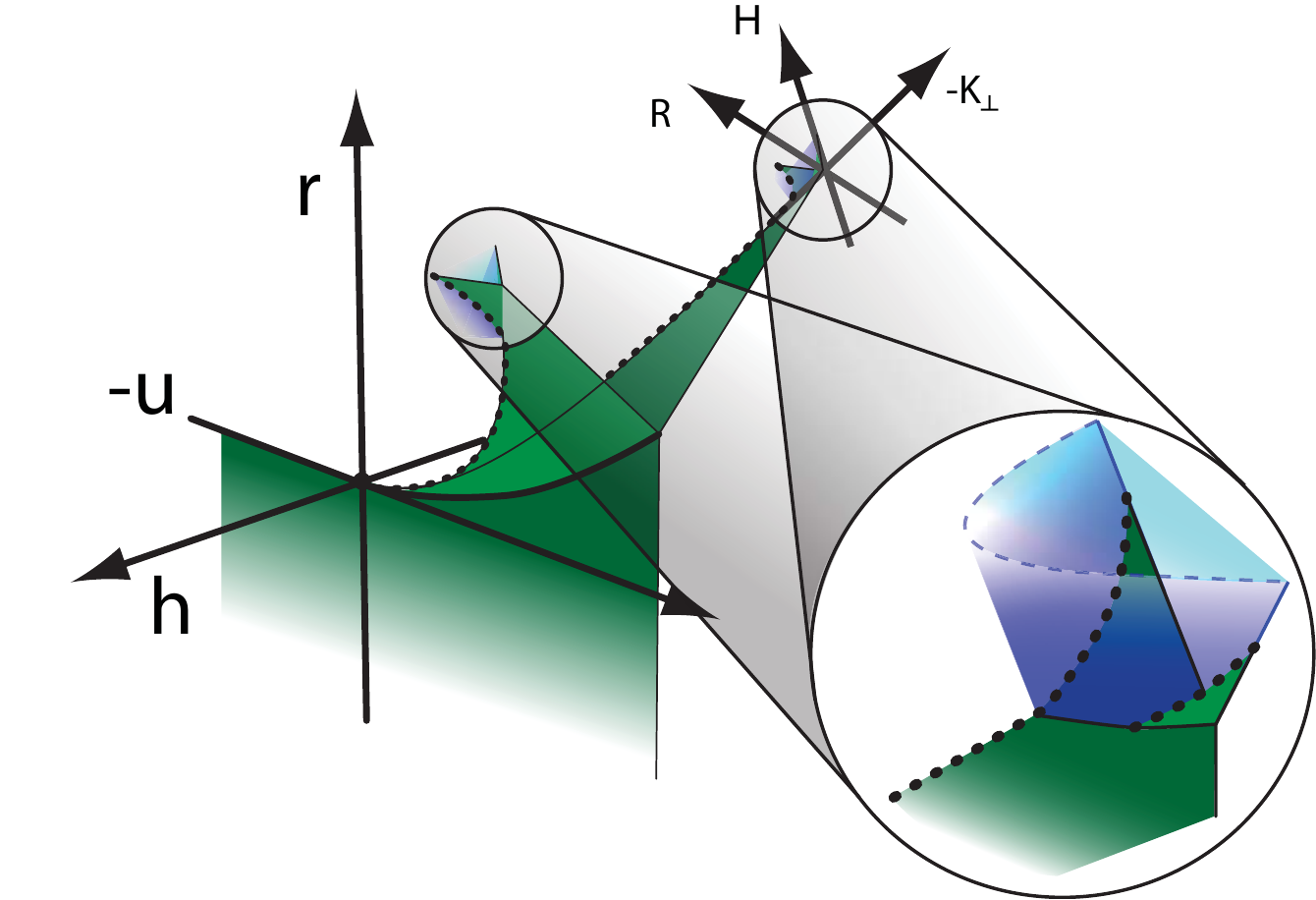}
\caption{\label{fig:FullPD} (Color online)  Sketch showing how the phase diagram of Fig.\ref{fig:PD} fits into the phase diagram for the metamagnetic system.  This consists of a rotated version of Fig.\ref{fig:PD} placed on the metamagnetic wing of Fig.\ref{fig:LPD}.  The light blue (pale grey) region is the inhomogeneous magnetic phase, the dark green (grey) sheet is the $h=0$ transition, the green ``wings'' are the metamagnetic transition.}
\end{figure}

Recalling that this phase diagram is constructed from an expansion about the line of critical endpoints, the full phase diagram for the metamagnetic
system is obtained by placing the bifurcated structure back into the context of the metamagnetic wing.  The inhomogeneous phase then appears as shown in Fig.\ref{fig:FullPD}. 

\section{Comparison with Experiment}

\subsection{Experimental results}

\begin{figure*}
\centering
\includegraphics[width=\textwidth]{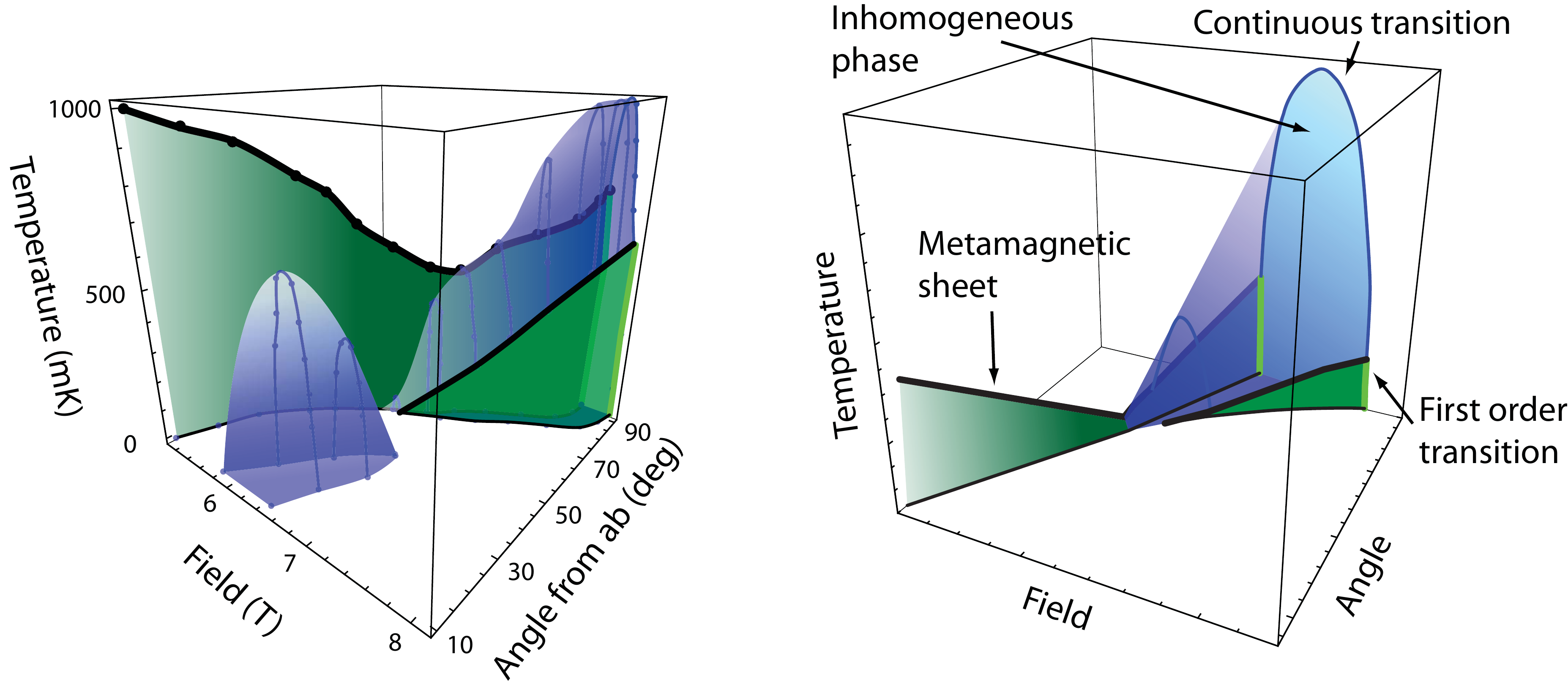}
\caption{\label{fig:EPD} (Color online)  a) The experimental phase diagram as inferred from in-plane transport properties.  Green (dark grey) planes correspond to abrupt changes in resistivity as a function of field. Blue (pale grey) shading indicates regions where the in-plane resistivity is anomalously high, becomes highly anisotropic with respect to the in-plane component of the field~\cite{Borzi07} and shows an anomalous temperature dependence: for currents in the direction of maximum resistivity, the resistivity  {\em decreases}  with increasing temperature.  The phase diagram obtained from magnetic susceptibility~\cite{Green05} shows the same  first order transitions as indicated here in green, but lacks the roof shown in blue.  b) Phase diagram from the Ginzburg-Landau theory rotated into experimental configuration.}
\end{figure*}

We now give a brief summary of the experimental results on Sr$_3$Ru$_2$O$_7$ before discussing the signatures of our proposed inhomogeneous state and how they compare.  The bilayered ruthenate Sr$_3$Ru$_2$O$_7$ shows a sequence of metamagnetic transitions~\cite{Grigera03a}. Early studies focussed on a line of metamagnetic critical end-points that could be tuned to a quantum critical point by adjusting the magnetic field strength and orientation~\cite{Grigera01}.  Subsequently, ultra-pure samples showed a bifurcation of this metamagnetic line upon approaching the putative quantum critical point~\cite{Grigera04,Green05} with a second line of critical end-points emerging from the zero-temperature plane as shown in Fig.\ref{fig:EPD}a.  This bifurcation is accompanied by a striking peak in resistivity~\cite{Grigera04} with curious, anisotropic dependence on the relative orientation of current, lattice and in-plane magnetic field~\cite{Borzi07}.  When current flows in the crystallographic direction most perpendicular to the in-plane field, the resistivity peak rapidly decreases as the field is moved away from the c-axis. When it is nearly parallel to the in-plane field, the peak persists. 

The bifurcating metamagnetic transitions are shown by green (dark grey) surfaces in Fig.\ref{fig:EPD} with the region of resistive anisotropy further delimited by the roof shown in blue (light grey). Indications of this roof were previously found with field along the c-axis~\cite{Grigera04} (including signatures in magnetostriction, magnetization and the temperature dependence of resistivity) Fig.\ref{fig:EPD} extends this roof in angle.  Similar features occur elsewhere in the phase diagram~\cite{Borzi07}, with further bifurcations apparent upon approaching the ab-plane.  These show a smaller resistance anomaly, but have the same characteristic anisotropy (the dome-shaped region in the foreground of Fig.\ref{fig:EPD}).

\subsection{Comparison of theory and experiment}

We now compare our model with results on Sr$_3$Ru$_2$O$_7$ and other materials.  Our model readily accommodates the behaviour of Sr$_3$Ru$_2$O$_7$.  

The phase diagram, Fig.\ref{fig:EPD}a, is obtained--- in the spirit of Ginzburg-Landau theory--- by interpreting $R$, $H$ and $K_\perp$ as linear functions of the experimental parameters $T$, $\theta$ and $h$. Fig.\ref{fig:EPD}b shows the result of such a transformation.  The resulting phase diagram has the same form as the experimental diagram.  In particular the sheet of continuous transitions into the inhomogeneous phase becomes a `roof'.  This roof encloses the anomalous phase and has been detected in several experimental probes~\cite{Grigera04,Rost}.  This roof is signaled by a qualitative change in the temperature dependence of resistivity and by a noticeable kink in the magnetization. We associate it with the continuous transition into the inhomogeneous phase found in the present theory. This roof previously presented a real puzzle as there is no obvious way to obtain it from a simple Landau theory for $\phi$, but it is nevertheless required to enclose the postulated broken symmetry phase in the bifurcated region.

The identification of the specific wavevector of modulation in Sr$_3$Ru$_2$O$_7$ would require a realistic model for the band structure of the material.  Quantum oscillation~\cite{Mercure} and ARPES~\cite{Tamai} results show that this band structure is extremely complex, making a full calculation extremely difficult.  However ARPES studies show a single band (named $\gamma_2$ in Ref.\onlinecite{Tamai}) with a peak in the DoS just below the Fermi energy~\cite{Tamai} which is implicated in the metamagnetic transition.  This is consistent with our single-band picture of a mechanism driven by a van Hove singularity.  It is therefore possible to work in a simplified picture containing only this band to capture the behaviour of Sr$_3$Ru$_2$O$_7$.  We expect that the wavevector of magnetic modulation will be associated with the nesting of these Fermi pockets.  We note that work has recently been carried out on a nematic distortion associated with this band~\cite{Kee}.

As the anomalous behaviour only appears in the cleanest samples, any mechanism that explains it must be sensitive to disorder.   Our mechanism  shows this sensitivity, since disorder smooths out features in the density of states.

\subsubsection{Band structure and angle dependence}

The natural parameters of our microscopic theory are field, temperature and band filling.  An additional mechanism is required to translate from filling to angle.  The most promising candidate is the field angle coupling to the underlying atomic orbitals and therefore modifying the band structure.  This mechanism has been investigated recently~\cite{Raghu,Kee}.

The Fermi surface of Sr$_3$Ru$_2$O$_7$ is made up predominantly from atomic orbitals of d$_{xy}$, d$_{zx}$ and d$_{yz}$ character.  The hybridization of these orbitals, together with bilayer splitting and backfolding due to structural distortion is responsible for the material's complex band structure~\cite{Tamai}.  By including spin-orbit coupling and both orbital and spin Zeeman effects the angular dependence of the Fermi surface can be calculated~\cite{Raghu}.

\begin{figure}
\centering
\includegraphics[width=3in]{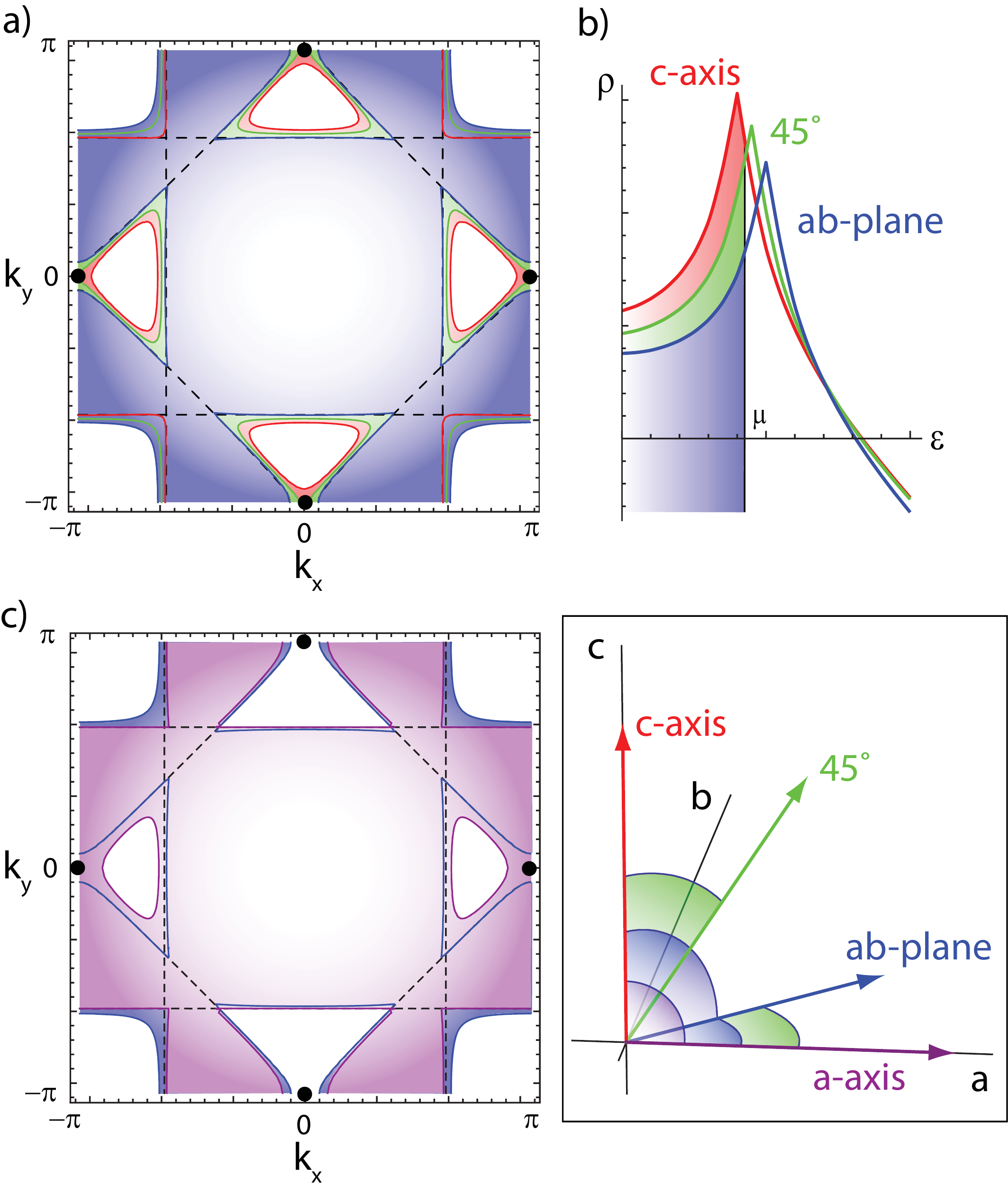}
\caption{\label{fig:angle} (Color online)  Angle dependence of simple model.  a)  Fermi surface for a minimal model of Sr$_3$Ru$_2$O$_7$.  Dotted lines show the unhybridized bands.  Colored lines indicate the Fermi surface of one spin species for a variety of angles:  Red, along c-axis.  Green, 45$^\circ$ from c.  Blue, ab-plane.  b) DoS for above angles.  The peak moves to lower energy as the angle from c increases.  c) An in-plane field component along one axis breaks the symmetry of the dispersion.  Inset, the various field angles illustrated in the Fermi surface plots.}
\end{figure}

We briefly present a minimal model to describe the effect of field angle.  Figure \ref{fig:angle} shows a Fermi surface made from three orbital bands, the quasi-1D d$_{zx}$ and d$_{yz}$ bands and a backfolded version of the 2D d$_{xy}$ band.  Angular dependence is included through orbital Zeeman coupling with the in-plane field component aligned at 45$^\circ$ to the a-axis.  As the field angle is changed from aligned with the c-axis to in-plane then the saddle point of the dispersion, and therefore the peak in the DoS, moves to lower energy.  In this way the critical field of the metamagnetic transition changes.  Work to fully include the angle in our model of spiral magnetization and to more accurately describe the Fermi surface of Sr$_3$Ru$_2$O$_7$ is ongoing.  

Spatially inhomogeneous magnetic structures lead inevitably to enhanced scattering in certain directions.  In order to fully explain the anisotropy, there must be a mechanism for an in-plane magnetic field to align the magnetic inhomogeneity. Our simple model does not contain such a mechanism. We suggest that its origin lies in the orbital effects of the in-plane field as described above. These modify the dispersion, breaking the symmetry between different orientations of the underlying helices when there is an in-plane field component, as seen in Fig. \ref{fig:angle}b).  When the sample is in the anomalous phase, there is significant magnetic inhomogeneity leading to enhanced resistivity.  With a magnetic field in the c-direction, the inhomogeneity does not break the crystal symmetry (at least macroscopically) and resistivity is isotropic.  This may be due to the formation of a spin crystal with the fourfold symmetry of the dispersion (see Fig.\ref{fig:angle}a), or through equal domains of each orientation.  As the field is rotated into the plane the distortion of the Fermi surface picks out a particular wavevector axis.  The magnetic inhomogeneity no longer preserves the lattice symmetry--- either through the formation of an anisotropic spin crystal or by a preponderance of domains of spin density waves of one orientation. This anisotropy is reflected in resistivity.  An explicit calculation of the resistivity is beyond the scope of this paper.  We may expect it to be proportional to the magnitude of $m_\perp$ which, as we have shown, goes continuously to zero at the `roof' of the anomalous phase.  This is in agreement with the measured resistivity anisotropy~\cite{Borzi07}.

\section{Discussion}

We conclude by summarizing our results and how they compare with experiment.  We briefly consider other theoretical proposals for the anomalous phase of Sr$_3$Ru$_2$O$_7$, and how these proposals could be distinguished.

We have shown that including the possibility of spatially modulated magnetization into the Stoner model allows for the formation of inhomogeneous states as an intermediate state in a metamagnetic transition.  This situation may already have been observed in Sr$_3$Ru$_2$O$_7$ where a bifurcation of the metamagnetic transition is accompanied by a region of anomalously high and anisotropic resistivity.  Our theory reproduces the topology of the experimentally observed phase diagram, including the bifurcation of the transition and a `roof' of continuous transitions enclosing the phase.

We have suggested a mechanism for the angle of the applied field to affect the properties of the system. Angle is known to play two roles: tuning the metamagnetic critical field to enable movement through the phase diagam at all, and aligning the resistivity anisotropy in the anomalous region.  We suggest that the effects of angle are likely to be due to orbital effects which modify the band structure~\cite{Raghu,Kee}.  In this scheme both spin-orbit and orbital-Zeeman coupling modify the Fermi surface based on the underlying orbitals.

There are several other theoretical proposals for modulated or anisotropic electronic states.  Here we briefly discuss the Dzyaloshinskii-Moriya interaction, quantum fluctuation corrections and nematic states.

In systems without a center of inversion symmetry spin-orbit interactions lead to a Dzyaloshinskii-Moriya interaction~\cite{Moriya,Dzyaloshinskii}.  This favours the formation of magnetic spirals~\cite{Bak} and possibly magnetic crystals~\cite{Binz06,Pfleiderer04,Rossler06}.  Here we are concerned with systems which do have inversion symmetry and so our results are not due to a Dzyaloshinskii-Moriya interaction.

Analysis of quantum fluctuation corrections to the theory of itinerant magnets suggests that they can induce metamagnetism and magnetic inhomogeneity~\cite{Belitz_1,Belitz_2,Belitz_3,Rech}.  Whether such effects can recover the phase behaviour of Sr$_3$Ru$_2$O$_7$ is unclear.  These effects may compete or act in parallel with the mechanism proposed here, their interaction with band structure induced features being as yet unexplored.  We expect that van Hove singularities are characterized by larger energy scales and dominate if present.

The anisotropic transport discussed here resembles that found in high Landau levels, where stripes of different filling factor break rotational and translational symmetry leading to smectic order. They may be aligned by small in-plane components of magnetic field leading to highly anisotropic resistivity~\cite{Roderich}. Melting of the stripe orientation may restore the translational symmetry giving rise to 
nematic order. Similarly, melting of the ordering of magnetic helices may lead to nematic order.  

Others have speculated that the anomalous phase may be a different type of nematic metal with a d-wave distortion of the Fermi surface~\cite{Fradkin07, Kee05}.  Recent work has centered on an orbital ordering which creates a similar distortion~\cite{Raghu,Wu,Kee}.  Although it invokes a different sector of the interaction, the model used to describe these distortions has many similarities to ours, with a very similar energetic drive.  The topology of the resulting phase diagram should be similar if extended in angle. The main experimental distinction is in the spatial modulation that we predict.  Spatial modulation of magnetization should show up as Bragg peaks in elastic neutron scattering in the anomalous region.  Unfortunately, such data do not exist. There are, however, pseudo-elastic data  outside of the anomalous region that are consistent with fluctuations that would freeze into the type of spin-crystals that we predict. These increase in intensity upon lowering temperature towards the anomalous phase and are consistent with the nesting vector of the $\gamma_2$ pockets~\cite{Antiferromagnet}.   Neutron scattering should ultimately reveal whether the order is of magnetic crystalline, or nematic type.

Our analysis is rather general and its results may have broader applicability.  {\it e.g.} NbFe$_2$~\cite{NbFe2} exhibits a peak in resistivity associated with the bifurcation of a metamagnetic transition and finite wavevector magnetic order and ZrZn$_2$~\cite{ZrZn2} may show similar features.

{\it Acknowledgments:} This work was supported by the Royal Society and the EPSRC under grant number EP/D036194/1. We are grateful to A.W. Rost, J.-F. Mercure, Gil Lonzarich and A.P. Mackenzie for insightful discussions.

\appendix*

\section{Understanding the Ginzburg-Landau theory}

The complexity of the Ginzburg-landau theory Eq.(\ref{GL2}) makes it hard to identify the role of each term.  In order to clarify this we break down the expansion and study the effect of each term one at a time.  As well as having explanatory value, this analysis will make explicit the process of calculating the phase diagram and reveal the fact that the smallness of ${\bf q}$ is not necessary for the phase reconstruction which we find.

The full free energy is broken into symmetric and antisymmetric segments with the cross-terms between $m_{||}$ and $m_\perp$ labelled $F_1$, $F_2$, and $F_3$:
\begin{widetext}
\begin{eqnarray}
\frac{\beta}{h \bar M} F_{\rm L}
&=&
\overbrace{\left( R \phi^2 + U \phi^4 + V \phi^6 - H \phi \right)
+
\left(
R_\perp + K_\perp {\bf q}^2 + L_\perp {\bf q}^4
\right) \phi_\perp^2
+
U_\perp
\phi_\perp^4
+ 
V_\perp \phi_\perp^6}^{F_0 \mbox{:  no cross-terms}}
\nonumber\\
&& 
+
\underbrace{U_1 \phi^2 \phi_\perp^2}_{F_1}
+
\underbrace{V_2 \phi^2 \phi_\perp^4}_{F_2}
+
\underbrace{
\left(
K_{2} {\bf q}^2\phi^2 + V_1 \phi^4 
\right) \phi_\perp^2
+ 
K_{3} {\bf q}^2 \phi_\perp^4}_{F_3}
+
\underbrace{K_{1} {\bf q}^2 \phi + T \phi^5 + \left( S_1 \phi + T_1 \phi^3 \right) \phi_\perp^2 + T_2 \phi \phi_\perp^4}_{F_a \mbox{:  antisymmetric}}.
\nonumber\\
\label{GL2A}
\end{eqnarray}
\end{widetext}
We will consider the effect of the various cross-terms on the phase diagram of the symmetric part before considering how the antisymmetic terms affect the phase diagram.

\subsection{Cross-terms: $U_1 \phi^2 \phi_\perp^2$}

\begin{figure}
\centerline{\includegraphics[width=3in]{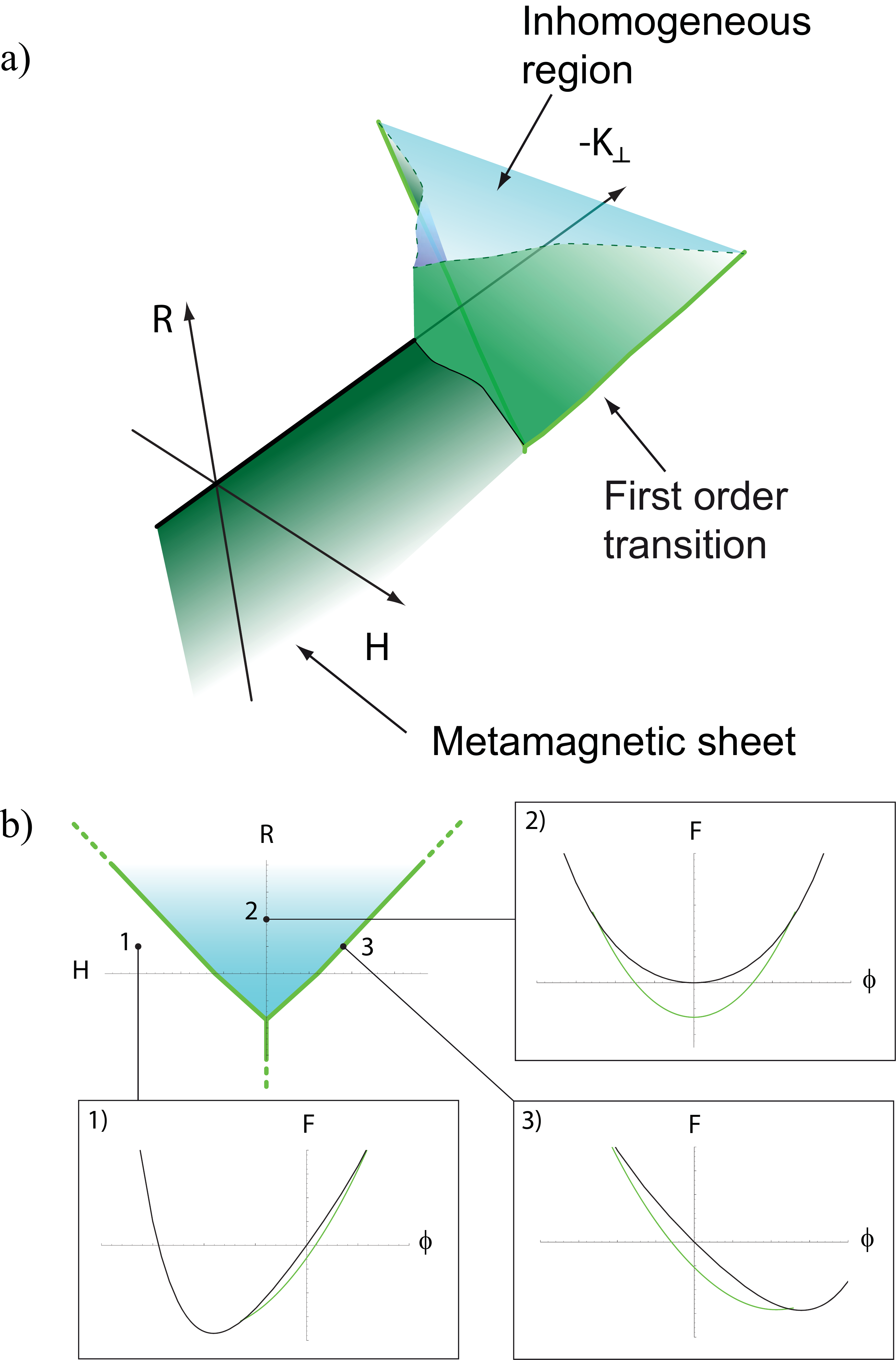}}
\caption{\label{fig:steps2} (Color online) a)  Phase diagram for a theory with a $U_1 \phi^2 {\phi_\perp}^2$ cross-term.  b)  A cut through the phase diagram at $K_\perp=-0.5$ with example free energy curves. Note the first order transition in panel 3 where two minima are degenerate.  [We take $L_\perp=0.1$]} 
\end{figure}

The minimal theory with interesting structure in the phase diagram includes the $F_1 = U_1 \phi^2 \phi_\perp^2$ cross-term that couples $\phi$ and $\phi_\perp$.
\begin{eqnarray}
\beta F(\phi, {\phi_\perp}, q)
&=&
h \bar M  \left(F_0+F_1\right).
\label{stepsGL2}
\end{eqnarray}
The phase diagram corresponding to this free energy is shown in Fig.\ref{fig:steps2}.  For small $K_\perp$ the longitudinal magnetization has a single metamagnetic transition.  At sufficiently negative $K_\perp$, the formation of inhomogeneous transverse magnetization becomes favourable.  The metamagnetic wing  splits into a Y shape, as shown in Fig.\ref{fig:steps2}.  The arms and leg of the Y are first order transitions in both $\phi$ and $\phi_\perp$ and extend to infinity.  Between the arms of this Y the transverse magnetization is inhomogeneous. 

Now let us discuss how this phase diagram follows from the free energy given in Eq.(\ref{stepsGL2}). The result of cross terms between longitudinal and transverse magnetization is that real, non-zero solutions for the transverse magnetization exist only in a restricted range of longitudinal magnetization and hence $H$. Our analysis proceeds by finding the optimum wavevector $\bar {\bf q}$ and optimum transverse magnetization $\bar \phi_\perp$ and substituting them back into the free energy to obtain an effective free energy $F_{\rm eff}(\phi)$. We present a graphical analysis of this free energy to give a feel for the structure of the phase diagram and show analytically why all of the phase transitions between homogeneous and inhomogeneous phases are first order in this simplified theory.

The optimum wavevector, found by minimizing over $q$ is $\bar q=\sqrt{-K_\perp/2L_\perp}$.  Substituting into Eq.(\ref{stepsGL2}) gives
\begin{eqnarray}
\frac{\beta}{h \bar M} F_{\rm eff}(\phi, {\phi_\perp})
&=&
R \phi^2 + U \phi^4 + V \phi^6 
 + \left( R_\perp' + U_1 \phi^2\right) \phi_\perp^2 
\nonumber\\
&& + U_\perp \phi_\perp^4 + V_\perp \phi_\perp^6.
\label{effGL2}
\end{eqnarray}
where $R_\perp'=R_\perp-K_\perp^2/(4L_\perp)$.  The optimum transverse magnetization calculated from this free energy is given by
\begin{equation}
\bar \phi_\perp^2 = \frac{-2U_\perp+\sqrt{4U_\perp^2-12V_\perp
    \left(R_\perp'+U_1 \phi^2\right)}}{6V_\perp} ,
\label{U1mperp}
\end{equation}

The effective free energy as a function of $\phi$,  $F_{\rm eff}(\phi)$, is obtained by substituting $\bar \phi_\perp^2$ from Eq.(\ref{U1mperp}) into $F_{\rm  eff}(\phi, {\phi_\perp})$ from Eq.(\ref{effGL2}). There are two subtleties in making this substitution. Firstly, in order that the free energy $F_{\rm eff}(\phi)$ be an expansion in powers of  $\phi$, we Taylor expand Eq.(\ref{U1mperp}) for $\bar \phi_\perp^2$ before substitution. Secondly, we must allow for the fact that $\bar \phi_\perp^2$ is only real and non-zero in certain regions. We account for this by introducing step functions, $\Theta$, that restrict the inhomogeneous contributions to the free energy to regions where $\bar \phi_\perp^2$ is real and positive. Substituting $\bar \phi_\perp^2$ into $F_{\rm  eff}(\phi)$ accounting for these considerations results in an effective free energy
\begin{eqnarray}
\frac{\beta}{h \bar M} F_{\rm eff}(\phi)
&=&
R \phi^2 + U \phi^4 + V \phi^6 - H \phi
\nonumber\\
&&+ \left(\alpha + \beta \phi^2+\gamma \phi^4+\delta \phi^6\right) \Theta
(\theta) \Theta(\bar \phi_\perp^2) 
\nonumber\\
\label{Feff}
\end{eqnarray}
where
\begin{eqnarray}
\alpha
&=&
\frac{2U_\perp^3-9R_\perp' U_\perp V_\perp-2 U_\perp^2 A +6 R_\perp' V_\perp A}{27 V_\perp^2},
\nonumber\\
\beta
&=&
\frac{U_1(U_\perp^2-3R_\perp' V_\perp -U_\perp A)}{3V_\perp A},
\nonumber\\
\gamma
&=&
-\frac{U_1^2}{4 A},
\nonumber\\
\delta
&=&
-\frac{U_1^3 V_\perp}{8 A^3},
\nonumber\\
A
&=&
\sqrt{U_\perp^2-3R_\perp'V_\perp},
\nonumber\\
\theta
&=&
4 U_\perp^2-12 V_\perp R_\perp'.
\label{endcoeff}
\end{eqnarray}
We now examine the transitions into the inhomogeneous phase. Fig.\ref{fig:steps2} shows the free energy plotted at various points on the phase diagram.  The black curve is the free energy with no contribution from inhomogeneity in the transverse magnetization.  The green (grey) curve shows the free energy with inhomogeneity in the transverse magnetization; {\it i.e} in the region where the step functions in Eq.(\ref{Feff}) are $1$.  We see that for low and high $H$ the global minima of the free energy lies on the homogeneous curve.  The free energy is minimised by a value of $\phi$ which corresponds to $\bar \phi_\perp=0$ and the system is in the homogeneous state.  For low $H$ we see that the global minimum lies on the inhomogeneous curve.  The inhomogeneous terms in the free energy have created an additional minimum of the free energy at low $\phi$.  For values of $H$ for which this is the absolute minimum of the free energy the system is in the inhomogeneous state. 

The nature of the transitions depend on the magnetizations corresponding to the minima of the free energy when they are degenerate.  At the transition point the global minimum of the free energy jumps discontinuously between a minimum in the homogeneous region and a minimum in the inhomogeneous region as $H$ is varied.  This results in a discontinuity in the optimum value of $\phi$ and a sudden jump to a non-zero value of $\phi_\perp$.  This is a first order transition in both longitudinal and transverse magnetization. We may construct a rigorous argument why the transitions are first order in $\phi$ and $\phi_\perp$ in the present simplified theory. For a transition between the homogeneous and inhomogeneous phases to be second order $\bar \phi_\perp^2$ must be zero at the transitions.  From Eq.(\ref{U1mperp}) we see that there is no real solution for $\phi$ when $\bar \phi_\perp^2=0$ and, therefore, that the transitions into the inhomogeneous phase are always first order.

\subsection{Cross-terms: adding the $V_2 \phi^2 \phi_\perp^4$ term}

\begin{figure}
\centerline{\includegraphics[width=3in]{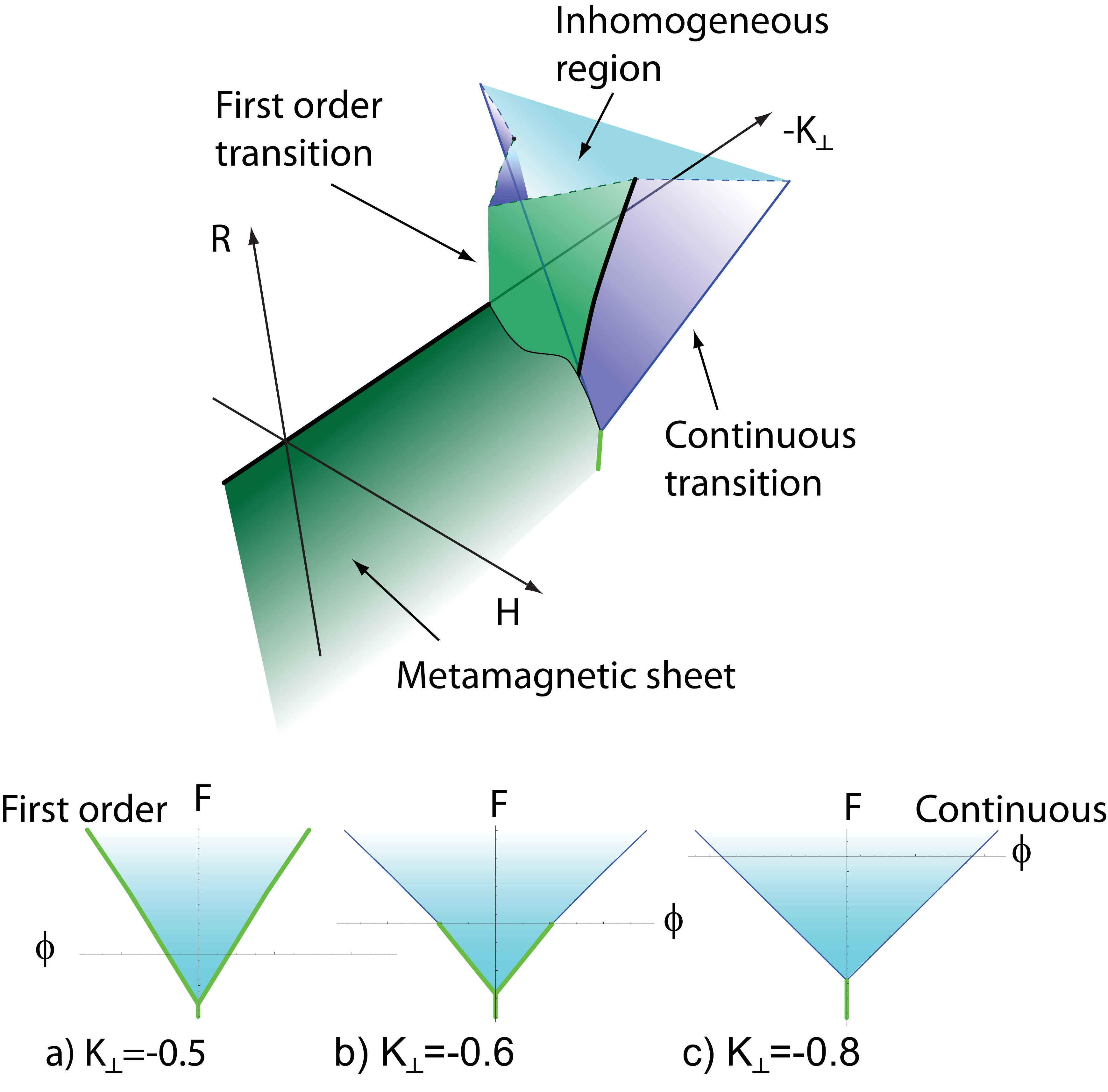}}
\caption{\label{fig:steps3} (Color online)  Phase diagram for a theory with $U_1 \phi^2 \phi_\perp^2$ and $V_2 \phi^2 {\phi_\perp^4}$ cross-terms.  Cuts show a) $K_\perp=-0.5$, all transitions first order.  b) $K_\perp=-0.6$, transitions change from first to second order at a critical point.  c)  $K_\perp=-0.8$, all transitions second order.  [We take $L_\perp=0.1$]} 
\end{figure}

Adding a further cross-term $V_2 \phi^2 \phi_\perp^4$ the free energy becomes
\begin{eqnarray}
\beta F(\phi, {\phi_\perp}, q)
&=&
h \bar M \left(F_0+F_1+F_2\right).
\label{stepsGL3}
\end{eqnarray}
The phase diagram now takes the form shown in Fig.\ref{fig:steps3}.  Many of the features of this phase diagram are the same as found in the preceding case.  The main difference between the phase diagram obtained from Eq.(\ref{stepsGL3}) (Fig.\ref{fig:steps3}) and that obtained from Eq.(\ref{stepsGL2}) (Fig.\ref{fig:steps2}) is that for Eq.(\ref{stepsGL3}) the transition may be either first order or continuous.  When the Y first appears the transition between homogeneous and inhomogeneous order is first order in both longitudinal and transverse magnetization. At more negative values of $K_\perp$, the transition becomes continuous above a critical value of $R$ --- as indicated by the change from green (thick) to blue (thin) lines in Fig.\ref{fig:steps3}.  Along these thin blue lines, the transition is second order in $\phi_\perp$ with inhomogeneous transverse magnetization appearing continuously from zero. The corresponding transition in longitudinal magnetization is also continuous but with a discontinuity in its gradient.

The structure of this phase diagram can be understood as before by optimising the free energy over $\phi_\perp$ and $q^2$ and substituting back their optimum values to obtain an effective free energy for the longitudinal magnetization $\phi$.  The optimum value of ${\phi_\perp}^2$ is 
\begin{eqnarray}
\bar \phi_\perp^2 &=& \frac{-2\left(U_\perp+V_2 \phi^2\right)}{6V_\perp}
\nonumber\\
&&+\frac{\sqrt{4\left(U_\perp+V_2 \phi^2\right)^2-12V_\perp
    \left(R_\perp'+U_1 \phi^2\right)}}{6V_\perp}.
\nonumber\\
\label{mp2}
\end{eqnarray}
Analysis of the expression for $\bar \phi_\perp^2$ allows us to deduce the order of transitions between the homogeneous and inhomogeneous 
regions of the phase diagram.  This is determined by the value of $R_\perp'=R_\perp-K_\perp^2/(4 L_\perp)$.  At a second order transition $\bar \phi_\perp$ must be zero.  This occurs when the $U_\perp+V_2 \phi^2$ term and the square root term of Eq.(\ref{mp2}) are zero.  These conditions are satisfied for real $\phi$ only when $R_\perp'<-\frac{1}{3}$.  This free energy gives first order transitions near to the point where the inhomogeneous phase first appears, which become second order as we move to more negative $K_\perp$. 

\subsection{All symmetric terms}

\begin{figure*}
\includegraphics[width=\textwidth]{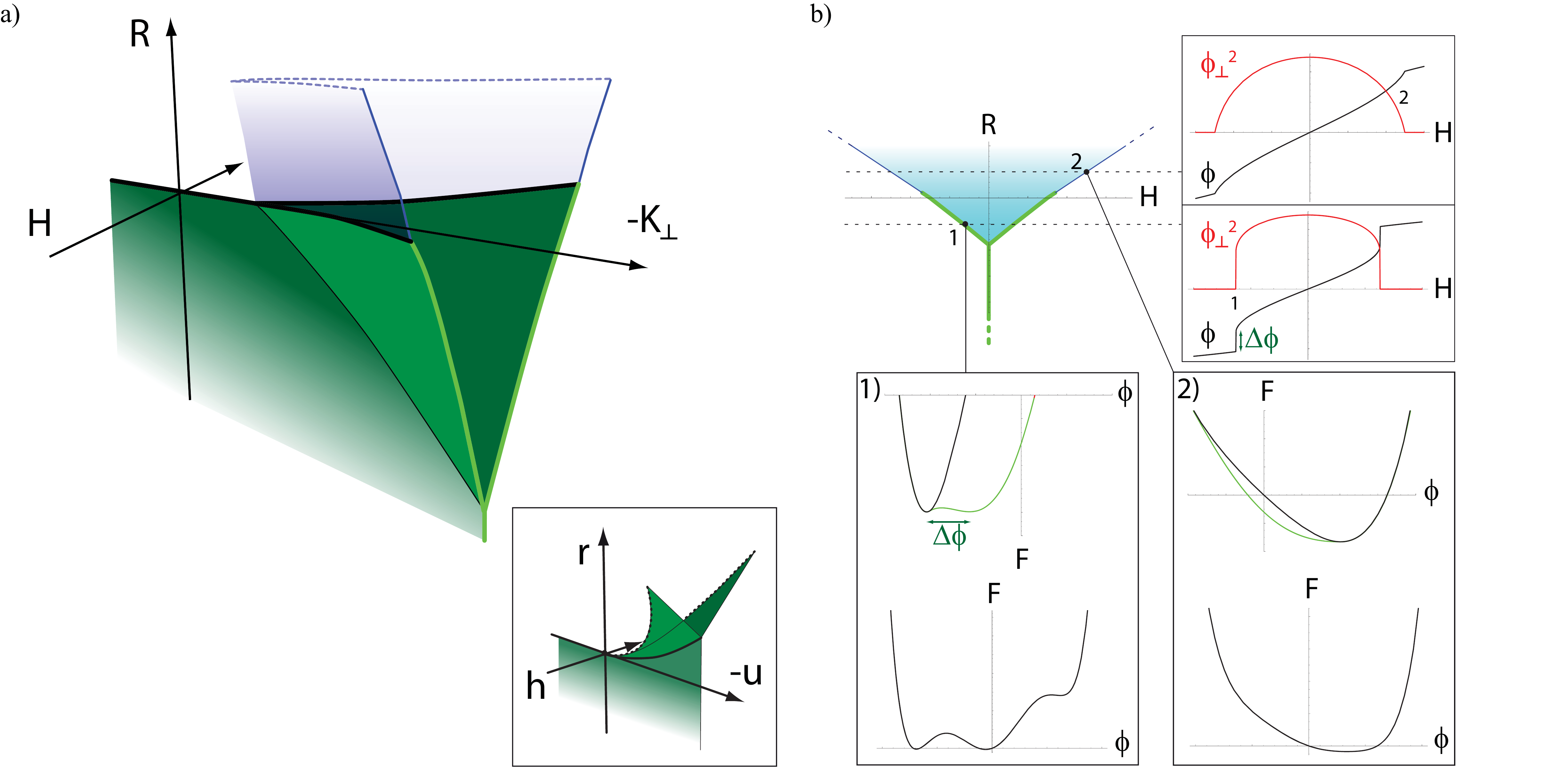}
\caption{\label{fig:symPD} (Color online) a)  Phase diagram for the symmetric theory with particular choices for $K_2$ and $K_3$.  Inset shows the tricritical point structure of a conventional Landau theory. b)  A cut through the phase diagram taken at $K_\perp=-1$ showing first order transitions below $R=0.2$ and second order transitions above.  Plots of the effective free energy show 1) a first order transition, and 2) a second order transition.  The top free energy curve shows how this occurs for the inhomogeneous theory and the bottom for a conventional theory.  Black curves are homogeneous terms only and green (grey) curves include inhomogeneous terms.  Also shown are magnetization plots as a function of $H$ showing both longitudinal (black) and transverse (red) magnetization.  The magnitude of the jump in $\phi$ at the first order transition ($\Delta \phi$)is related to the spacing of minima in the free energy.  [We take $L_\perp=0.1$, $K_2=0.3$ and $K_3=0.2$]. } 
\end{figure*}

We now include the remaining symmetric terms from the free energy Eq.(\ref{GL2}); $V_1 \phi^4 \phi_\perp^2$, $K_2 q^2 \phi^2 \phi_\perp^2$ and $K_3 q^2 \phi_\perp^4$.  The free energy is given by
\begin{eqnarray}
\beta F(\phi, {\phi_\perp}, q) 
&=& 
h \bar M \left(F_0+F_1+F_2+F_3\right).
\label{GLs}
\end{eqnarray}
The resulting phase diagram is shown in Fig.\ref{fig:symPD}.  The major modification from previous cases is the location of the line of critical points where the transition between homogeneous and inhomogeneous order becomes first order. These lines converge upon the parent line of metamagnetic critical end-points to form a tricritical point structure ---although this is of an unusual type. As we will see in more detail below, the most important of the additional terms in driving this restructuring of the phase diagram is the ${\bf q}^2 \phi_\perp^4$ term. 

As in the preceding analysis, the phase diagram is obtained by considering an effective free energy for the longitudinal magnetization, $\beta F_{\rm eff}(\phi)$. The first step in deriving this effective free energy is to optimize over ${\bf q}$. Because of the additional, momentum-dependent cross terms between $\phi$ and ${\bf \phi}_\perp$, $\bar {\bf q}$ is not constant, but depends upon $\phi$;
\begin{eqnarray}
\bar q^2=-\frac{K_\perp+K_2 \phi^2+K_3 \phi_\perp^2}{2L_\perp}.
\label{OptQ}
\end{eqnarray}  
The optimum value of ${\phi_\perp}^2$ is given by
\begin{eqnarray}
{\bf \bar \phi_\perp}^2&=&\frac{1}{6 \left( V_\perp'\right)}
\left(-2\left(U_\perp'+V_2' \phi^2\right)+\right. 
\nonumber\\
&&\left.\sqrt{4\left(U_\perp'+V_2'
      \phi^2\right)^2-12V_\perp'\left(R_\perp'+U_1' \phi^2+V_1'
      \phi^4\right)}\right) .
\nonumber\\
\label{OptPhi}
\end{eqnarray}
where the dashed coefficients are simple functions of the undashed coefficients in the free energy.

The phase diagram shown in Fig.\ref{fig:symPD}a is similar to that of a conventional tricritical point (inset to Fig.\ref{fig:symPD}a. Indeed, the topology of the first order transitions in the phase diagrams  --- indicated by the green (dark grey) surfaces in Fig.\ref{fig:symPD} --- is identical. There are, however,  important differences due to the phase of inhomogeneous transverse magnetization which has produced the bifurcation.

A cut through the phase diagram at negative $K_\perp$, as shown in Fig.\ref{fig:symPD}b, consists of a Y shape with arms and leg that extend to infinity.  The arms begin by describing a first order transition in both longitudinal and transverse magnetization --- indicated in green (thick line). At a critical value of $R$, the order of the transition changes --- in a conventional metamagnet, the arm would stop here at a critical end point. Along the blue (thin) lines in Fig.\ref{fig:symPD}b and over the entire blue (pale grey) surface in Fig.\ref{fig:symPD}a a second order transition to non-zero transverse inhomogeneous magnetization is accompanied by a kink in the longitudinal magnetization. This latter feature has no analogue near the conventional tricritical point. Moving towards positive $K_\perp$, the point at which the transition becomes continuous gets closer to the junction of the Y until they coincide at the tricritical point.

A comparison of free energy curves is made for typical points along the metamagnetic wing in Fig.\ref{fig:symPD}b. A first order transition occurs when local minima of $F_{\rm eff}(\phi)$ have the same free energy.  For the present theory one of the minima is in the inhomogeneous phase (shown in green [grey]) and the other in the homogeneous phase (shown in black).  This leads to a jump in longitudinal magnetization, $\Delta \phi$, and also a jump to non-zero inhomogeneous transverse magnetization. For the conventional case there is no inhomogeneous phase, but the transition remains a jump between two homogeneous minima. As we move along the transition line in the direction of increasing $R$ the size of the jump in magnetization decreases. Along the thin blue line, the minimum of the free energy swaps continuously between the homogeneous (black) and inhomogeneous (green [grey]) curves. Near the conventional tricritical point there is no signature in the free energy along this line.

This structure of the bifurcated metamagnetic wings and of the crucial role of the $K_3 {\bf q}^2 \phi_\perp^4$ term can be appreciated from an analysis of Eq.(\ref{OptPhi}).  This is simplified by restricting the analysis to the vicinity of the tricritical point so that $\phi$- and $\phi_\perp$-dependent terms can be neglected in Eq.(\ref{OptQ}) and Eq.(\ref{OptPhi}). In this limit, the renormalized $\phi_\perp^4$ coefficient is given by $U'_\perp = U_\perp+K_3 \bar q^2$. In fact, the optimum wavevector is only weakly dependent upon $\phi$ and this renormalization is the dominant effect of the $K_3 {\bf q}^2 \phi_\perp^4$ term throughout the inhomogeneous regime. At a second order transition, $\bar \phi_\perp$ must be zero. Near to the tricritical point this leads to the condition 
\begin{eqnarray}
-2U_\perp'+\sqrt{4U_\perp'^2-12V_\perp R_\perp'}=0.
\end{eqnarray}
If $U_\perp'$ (recall that $U_\perp=-1/8$) is negative then this cannot be satisfied. If the renormalization of $U_\perp$ due to the the $K_3 {\bf q}^2 \phi_\perp^4$ term is large enough then $U_\perp'$ becomes positive. The condition for a continuous transition becomes $R_\perp'<0$, which is by definition satisfied in the inhomogeneous region, thus implying that the transition into the inhomogeneous region is continuous at the tricritical point.

\subsection{Adding antisymmetric terms}

\begin{figure}
\centerline{\includegraphics[width=3in]{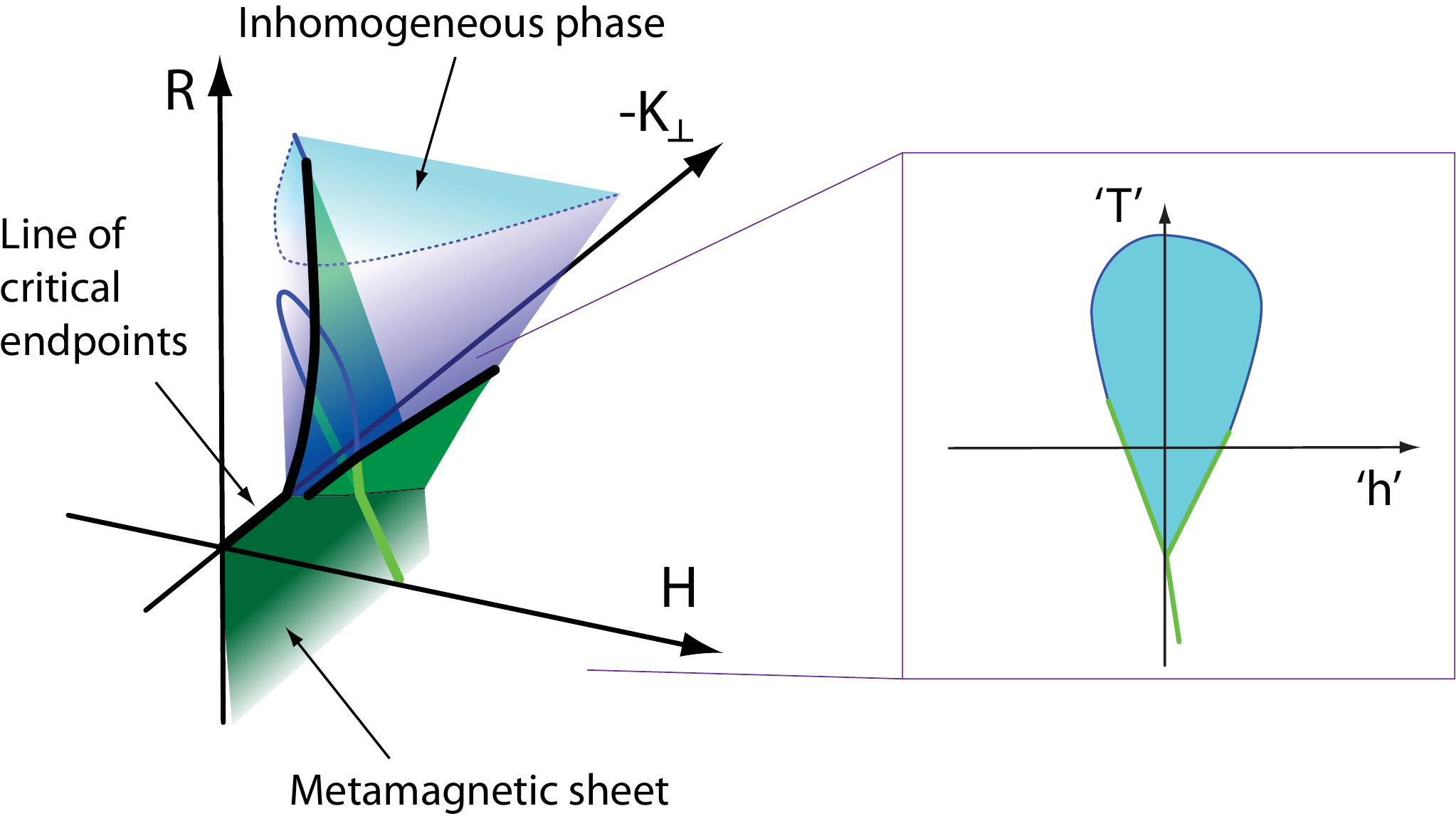}}
\caption{\label{fig:PhaseDiagramAgain} (Color online) Phase diagram for the Ginzburg-Landau theory as in Fig.\ref{fig:PD}.  Green (dark grey) sheets represent first-order transitions in $\phi$.  Blue (pale grey) sheets represent continuous transitions into the inhomogeneous phase.  $K_{\perp}$ represents movement along the metamagnetic wing.  $H$ moves in a direction perpendicular to the wing.  Taking a cut at an angle to the $R$-axis shows how the inhomogeneous region becomes finite.  Solid blue (thin) and green (thick) lines show cuts at constant $K_\perp$.} 
\end{figure}

When we add antisymmetric terms the tricritical point becomes symmetry broken, or dislocated, as shown in Fig.\ref{fig:PhaseDiagramAgain}. A conventional Landau theory of a dislocated tricritical point was previously proposed by Green {\it et al}~\cite{Green05}. Whilst the latter theory captured the phase diagram obtained from longitudinal magnetic susceptibility, it could not accommodate the `roof' over the region of anomalous transport found experimentally~\cite{Grigera04} (see Fig.\ref{fig:EPD} of the main text).  The blue (pale grey) surface in Fig.\ref{fig:symPD} correctly reproduces the features of the roof as seen in the cut through Fig.\ref{fig:PhaseDiagramAgain}. The phase diagram after suitable reorientation in the spirit of Landau theory is shown in Fig.\ref{fig:EPD}b.

Finally, we comment upon the wavevector of the inhomogeneous order. So far we have assumed that this is small and performed a standard Ginzburg-Landau expansion in powers of ${\bf q}$.  In fact, it is not necessary that $\bar q$ be small in order to obtain the phase reconstruction discussed here. As indicated in our discussion of the role of the $K_3{\bf q}^2 \phi_\perp^4$ term above, the optimum wavevector stays largely constant through the inhomogeneous phase. Its role is mainly to renormalize various homogeneous coefficients in the free energy. It is not necessary that $\bar q$ be small in order to fulfill this role. The only requirement is that the inhomogeneous order become favourable at some point along the line of metamagnetic critical end-points upon moving away from the parent tricritical point. As we saw in the microscopic analysis, it is plausible that inhomogeneity occurs at either small or large wave-vectors depending upon the details of the electronic dispersion.

\end{document}